\documentclass[reprint,superscriptaddress,amsmath,amssymb,aps,pra,longbibliography]{revtex4-1}

\ifx\pdfoutput\@undefined\usepackage[usenames,dvips]{color}
\else\usepackage[usenames,dvipsnames]{color}
\usepackage{graphicx}
\usepackage{bm}
\usepackage{amssymb}
\usepackage{hyperref}
\usepackage{color}

\usepackage{amsmath}
\usepackage{eufrak}

\usepackage{epigraph}
\usepackage{lipsum}

\usepackage[T1]{fontenc}
\usepackage[latin1]{inputenc}

\newcommand{\be}[0]{\begin{equation}}
\newcommand{\ee}[0]{\end{equation}}
\newcommand{\bea}[0]{\begin{align}}
\newcommand{\eea}[0]{\end{align}}

\newcommand{\br}[0]{{\bf r}}
\newcommand{\bu}[0]{\bar{\bf u}}
\newcommand{\bv}[0]{\bar{\bf v}}
\newcommand{\bw}[0]{\bar{\bf w}}

\newcommand{\ua}[0]{\bar{\bf a}}
\newcommand{\ub}[0]{\bar{\bf b}}

\newcommand{\deriv}[0]{\partial_{\tau}}

\newcommand{\CMRD}{\textcolor{black}}
\newcommand{\CKB}{\textcolor{black}}
\newcommand{\KB}{\textcolor{black}}
\newcommand{\KBII}{\textcolor{black}}

\usepackage{graphicx}
\usepackage{color}

\begin{document}

\title{\Large Geometric phases in 2D and 3D polarized fields: \\ 
geometrical, dynamical, and topological aspects}

\author{\large Konstantin Y. Bliokh}
\affiliation{\normalsize RIKEN Cluster for Pioneering Research, Wako-shi, Saitama 351-0198, Japan}
\affiliation{\normalsize Nonlinear Physics Centre, RSPE, The Australian National University, Canberra, Australia}
\author{\large Miguel A. Alonso}
\affiliation{\normalsize Aix Marseille Univ, CNRS, Centrale Marseille, Institut Fresnel, UMR 7249, 13397 Marseille CEDEX 20, France}
\affiliation{\normalsize The Institute of Optics, University of Rochester, Rochester, New York 14627, USA}
\author{\large Mark R. Dennis}
\affiliation{\normalsize School of Physics and Astronomy, University of Birmingham, Birmingham B15 2TT, UK}


\begin{abstract}
Geometric \CMRD{phases are} a universal concept that underpins numerous phenomena involving multi-component wave fields. These polarization-dependent phases are inherent in interference effects, spin-orbit interaction phenomena, and topological properties of vector wave fields.
Geometric phases \CMRD{have been} thoroughly studied in two-component fields, such as two-level quantum systems or paraxial optical waves\CMRD{. However,} their description for fields with three or more components, such as generic nonparaxial optical fields routinely used in modern nano-optics, constitutes a nontrivial problem. 
Here we describe geometric, dynamical, and total phases calculated along a closed spatial contour in a multi-component complex field, with particular emphasis on 2D (paraxial) and 3D (nonparaxial) optical fields. We present several equivalent approaches: (i) an algebraic formalism, universal for any multi-component field\CMRD{;} (ii) a dynamical approach using the Coriolis coupling between the spin angular momentum and reference-frame rotations\CMRD{;} and (iii) a geometric representation, which unifies the Pancharatnam-Berry phase for the 2D polarization on the Poincar\'e sphere and the Majorana-sphere representation for the 3D polarized fields.  
Most importantly, we reveal close connections between geometric phases, angular-momentum properties of the field, and topological properties of polarization singularities in 2D and 3D fields, such as C-points and polarization M\"obius strips.

\end{abstract}

\maketitle

{\epigraph{\CKB{There is geometry in the humming of the strings, there is music in the spacing of the spheres.}}{Pythagoras}
\epigraph{\CKB{Geometry is not true, it is advantageous.}}{H. Poincar\'e}}

\section{\normalsize Introduction}

Geometric phases, \CMRD{recognized} as a universal phenomenon 35 years ago by Michael Berry \cite{Berry1984,Shapere_book}, play \CMRD{a fundamental} role in the interference of vector waves, \CMRD{especially} monochromatic optical fields \KB{\cite{Vinitskii1990,Bhandari1997,Ben-Aryeh2004,Malykin2004,Bliokh2009,Karimi2019}}. 
They are therefore ubiquitous in several areas of modern optics including polarization manipulations \cite{Bhandari1997,Hasman2005,Marrucci2011,Bliokh2015NP}, singular optics \cite{Nye_book,Soskin2001,Dennis2009}, and the angular momentum (AM) of light \cite{Allen_book,Andrews_book,Molina2007,Franke2008,Bliokh2015PR}. 
Geometric phases are \CMRD{crucial as both a fundamental concept and a} practical tool underlying optical elements, such as ``metasurfaces'' and ``q-plates'' \cite{Hasman2005,Marrucci2011,Bliokh2015NP,Capasso2014,Chen2016,
Xiao2017,Maguid2016,Maguid2017,Alonso2019,Rubano2019}.

The polarization properties of paraxial optical fields have been thoroughly studied. For these fields, the electric field vector is essentially constrained to the plane orthogonal to the propagation direction. 
Such 2D fields are characterized \CMRD{topologically by {\it polarization singularities}}: C-points (polarization vortices) and L-lines \cite{Nye1983,Dennis2009}.
However, with the rapid development of nano-optics and photonics, more attention has been placed on \CMRD{inhomogeneous 3D fields whose polarization may be nonparaxial}. 
In particular, \CMRD{structured fields can give rise to knotted 3D singularities \cite{Leach2004,Dennis2010,Laroque2018,Maucher2019,Sugic2019}} and polarization M\"{o}bius strips \cite{Freund2010,Freund2010II,Dennis2011,Bauer2015,Galvez2017,Garcia2017,Bauer2019}, as well as to the coupling of the spin and orbital AM \cite{Dogariu2006,Zhao2007,Bomzon2007,Bliokh2010,Bliokh2011,Bliokh2015NP,Picardi2018} where geometric phases play a key role.

Despite enormous progress in the investigations of nonparaxial 3D vector fields and their properties, the self-consistent characterization of {\it geometric and dynamical phases} in such fields remains somewhat elusive. 
Indeed, there are two main types of geometric phase known in polarization optics: (i) the Pancharatnam-Berry (PB) phase \cite{Pancharatnam,Berry1987,Bhandary1988,Simon1988} and (ii) the spin-redirection (or Bortolotti-Rytov-Vladimirskii-Berry) phase \cite{Bortolotti,Rytov,Vladimirskii,Ross1984,Chiao1986,Tomita1986,Berry1987_II}. 
The PB phase appears for paraxial 2D fields with evolving ${\rm SU(2)}$ transverse polarization but propagating in the same direction; it is \CMRD{represented geometrically} on the ${\rm S^2}$ Poincar\'{e}-Bloch sphere. 
In turn, the spin-redirection phase involves waves with constant circular polarization (helicity) but whose propagation direction evolves; it is geometrically represented on the unit ${\rm S^2}$ sphere of propagation directions (wavevectors). Thus, both of these phases are ${\rm SU(2)} \simeq {\rm SO(3)}$ geometric phases, typical for spin-1/2 quantum systems with {\it two-component} wavefunctions \cite{Berry1984}. However, generic 3D polarized waves have more degrees of freedom, and their properties should rather correspond to spin-1 waves with {\it three-component} wavefunctions. 

There have been only a few studies of geometric phases in general three-level and spin-1 quantum systems \cite{Bouchiat1988,Khanna1997}, as well as attempts to generalize the PB and spin-redirection geometric phases in optics into one unified phase \cite{Bhandary1989}. These attempts resulted in the geometric {\it Majorana sphere} formalism developed by Hannay \cite{Hannay1998,Hannay1998_II} and in the dynamical description in terms of the {\it Coriolis (or angular-Doppler) effect} for waves carrying spin AM \cite{Bliokh2008}. Still, these prior works do not provide a clear, unambiguous answer to the basic question ``{\it what are the geometric and dynamical phase increments along a given spatial contour in a generic 3D complex vector field?}''. Another thought-provoking problem is the {\it relations between the geometric phases, polarization singularities, and their topological properties in 2D and 3D fields}. We aim to address these questions in the present work.

The paper is organized as follows. We first introduce self-consistent definitions of the geometric and dynamical phases along a given spatial contour for an {\it arbitrary multi-component complex field} (Section~\ref{basic_idea}). The dynamical phase is always {\it quantized} (i.e., it equals an integer times $\pi$) for closed contours. 
Then, we apply this approach to 2D polarized fields (Section~\ref{Pancharatnam}) with generic polarization singularities: {\it C-points}, which are characterized by \CMRD{a} {\it topological $\mathbb{Z}$ number}. 
In this case, the geometric phase is the PB phase represented on the Poincar\'{e} sphere, but we show that its behavior exhibits \CMRD{discontinuities}, i.e., ``$\pi$ times the topological index'' jumps when smooth deformations of the contour cross a C-point. 

Next, we consider geometric and dynamical phases in generic 3D polarized fields (Section~\ref{Sec-General3D}). 
The 3D geometric phase, unifying the PB and spin-redirection phases, allows both dynamical Coriolis-effect description \cite{Bliokh2008}   (Section~\ref{Sec-Coriolis}) and a geometric representation on a Majorana-type sphere for spin-1 \cite{Hannay1998,Hannay1998_II} (Section~\ref{Sec-Poincarana}). Importantly, we introduce a new Majorana-like representation that allows a more natural geometric-phase interpretation and closer correspondence to the 2D Poincar\'{e}-sphere formalism. We call this construction the ``{\it Poincarana sphere}''. Remarkably, the topological properties of the contour on the Poincarana/Majorana sphere are intimately related to the topological properties of the 3D polarization distribution along the corresponding spatial contour. In this case, the {\it topological $\mathbb{Z}_2$ number} distinguishes the {\it M\"{o}bius and non-M\"{o}bius} polarization strips, and the geometric phase exhibits $\pi$ jumps at transitions between these two cases.

In Section~\ref{Sec-Other}, we consider the close relation of the geometric, dynamical, and total phases to the {\it spin, total, and orbital angular momenta} in cylindrical 3D optical fields, as well as \CMRD{extending them} to the case of electric and magnetic  fields {\it in optical media}.
Section~\ref{Discussion} concludes the paper and discusses \CMRD{the} physical implications of the \CMRD{overall} results presented in this work.

\section{\normalsize General approach and \\ paraxial 2D case
}
\subsection{\normalsize Geometric and dynamical phases in multicomponent complex fields}
\label{basic_idea}

\begin{figure}
\centering
\includegraphics[width=0.8\linewidth]{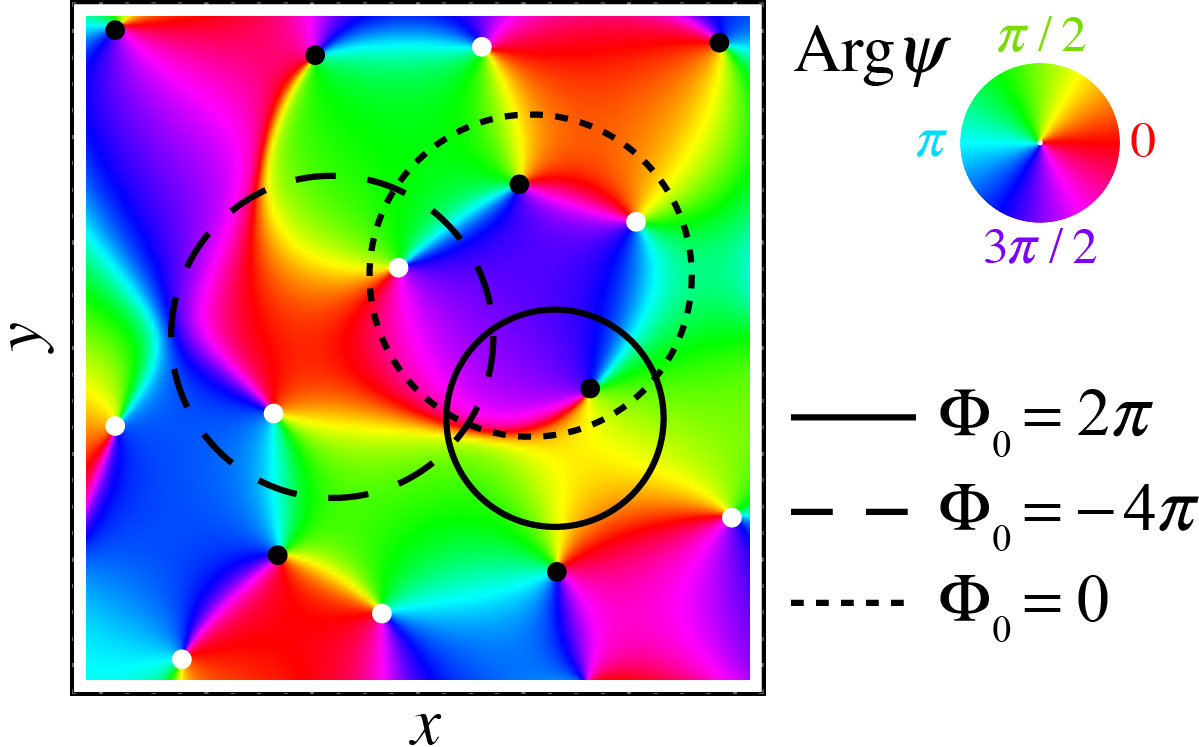}
\caption{An example of the 2D distribution of the colour-coded phase $\alpha={\rm Arg}\,\psi$ of a random scalar field $\psi({\bf r})$ \cite{Soskin2001,Dennis2009}. 
Phase singularities (vortices) of positive and negative unit charges ($N=\pm 1$ for the phases growing and decreasing upon \CMRD{counterclockwise} motion around the singularity) are marked by the black and white \CMRD{dots respectively}.
Three contours and the corresponding phases $\Phi_0$, Eq.~(\ref{scalar_phase}), indicating the net topological charges enclosed by the contours, $N=1$, $N=-2$, and $N=0$, are shown. A counterclockwise motion along circular contours in the $(x,y)$-plane is 
\CMRD{always assumed}.}
\label{Fig_scalar}
\end{figure}

The simplest case of a complex \CMRD{wavefield}, a {\it scalar} \CMRD{wavefield}, is described by a complex coordinate-dependent wavefunction $\psi(\bf{r})$, where \CMRD{we only} consider monochromatic fields and omit the time-dependence $e^{-i\omega t}$ factor. 
Its local phase $\alpha = {\rm Arg}\, \psi$ is defined via $\psi ( {\bf{r}} ) = \left| {\psi ( {\bf{r}} )} \right|{e^{i\alpha ( {\bf{r}})}}$. The gradient of this phase determines the {\it local wavevector} (or normalized canonical momentum density) ${{\bf{k}}_{\rm loc}} = \nabla \alpha$ \cite{Berry2009}, whose integral along a closed contour is {\it quantized} as an integer times $2\pi$:
\begin{equation}
\label{scalar_phase}
\Phi_{0}  = {\mathop{\rm Im}\nolimits} \oint {\frac{{{\psi ^*}\left( \nabla  \right)\psi }}{{{{\left| \psi  \right|}^2}}} \cdot d{\bf{r}}}  = \oint {\nabla \alpha  \cdot d{\bf{r}}}  = 2\pi N~.
\end{equation}
The integer $N$ counts the number of {\it phase singularities} or {\it vortices} (i.e.,  zeros of the field, where $|\psi|=0$ and $\alpha$ is undetermined) enclosed by the contour and weighted by their topological strengths \cite{Soskin2001,Dennis2009}. In other words, $N$ provides the total {\it topological charge} inside the contour, Fig.~\ref{Fig_scalar}. The phase singularities are generically points in 2D space, ${\bf r}=(x,y)$, and lines in 3D space, ${\bf r}=(x,y,z)$. In the latter case, one should consider a 2D surface spanning the contour and count point singularities on this surface, because the topological charge is well-defined only in the 2D case. Upon continuous deformations of the contour of integration, the phase (\ref{scalar_phase}) experiences jumps each time that the contour crosses singularities of the field $\psi$. Note also that only vortices with minimal charges $N=\pm 1$ are generic (see Fig.~\ref{Fig_scalar}); the higher-order singularities represent {\it degenerate} cases, such as in cylindrically symmetric higher-order vortex beams \cite{Allen_book,Andrews_book,Molina2007,Franke2008,Bliokh2015PR}.

\CMRD{Throughout this work, we consider the `phase' associated exclusively with closed contours.  When the integrand of the closed contour integral is the gradient of a phase function, as in (\ref{scalar_phase}) with $\alpha$, then the result of the contour integral counts the topological weighting of the phase singularities inside the contour.
When the integrand is not explicitly a gradient, then the phase around the contour changes typically with any deformation, i.e., the phase is purely geometric rather than topological.}

The case of {\it vector} \CMRD{wavefields} is more involved. In the main part of this work we consider the complex electric-field amplitude ${\bf E}(\bf{r})$ of a monochromatic optical field, although the approach of this subsection \CMRD{applies} to {\it any complex multi-component field} ${\bf E}=(E_1,...,E_n)$. 
On the one hand, a straightforward extension of the above scalar-field equations allows the introduction of the local wavevector (canonical momentum density) as the weighted average of the local wavevectors for each field component: ${\bf k}_{\rm loc} = {\rm Im} \left[ {\bf  E}^*\! \cdot \left( \nabla  \right) {\bf E} \right] / \left| {\bf E} \right|^2$, where $\left[ {\bf  E}^*\! \cdot \left( \nabla  \right) {\bf E} \right]_i \equiv \sum_{j=1}^{n}{E}^{*}_{j}  \nabla_i {E}_j$ \cite{Berry2009,Berry2013,Bliokh2013,Antognozzi2016}. However, unlike Eq.~(\ref{scalar_phase}), its integral along a closed contour is {\it not quantized}:
\begin{equation}
\label{total_phase}
\Phi  = {\rm Im} \oint {\frac{{{{\bf{E}}^*}\! \cdot \left( \nabla  \right){\bf{E}}}}{{{{\left| {\bf{E}} \right|}^2}}} \cdot d{\bf{r}}}  \ne 2\pi N~.
\end{equation}
\CMRD{Typically, this phase changes continuously upon continuous deformations of the contour of integration and does not have singularities, because it is not the gradient of a single phase function.}
\CMRD{One can nevertheless} reduce the vector-field problem to the scalar-field one by introducing the complex scalar field $\Psi  = {\bf{E}} \cdot {\bf{E}}$ \cite{BerryDennis2001}. 
The phase \CMRD{of $\Psi$} can be calculated according to Eq.~(\ref{scalar_phase}), and, since this field is quadratic \CMRD{in the components of} the original field ${\bf E}$, the phase \KBII{in the original field} acquires a factor of 1/2: 
\begin{equation}
\label{dynamical_phase}
{\Phi _{\rm D}} \! = \frac{1}{2}{\rm Im} \oint {\frac{{{\Psi ^*}\left( \nabla  \right)\Psi }}{{{{\left| \Psi  \right|}^2}}} \cdot d{\bf{r}}}  = \frac{1}{2}\oint {\nabla {\rm Arg} \Psi  \cdot d{\bf{r}}}  = \pi N_{\rm D},
\end{equation}
where $N_{\rm D}$ is an integer. Thus, this phase is quantized as a {\it half-integer} times $2\pi$, and generically it has singularities as any scalar-field phase.

Importantly, one can associate the phases (\ref{total_phase}) and (\ref{dynamical_phase}) with the {\it total} and {\it dynamical} phases, respectively, in a \CMRD{multi}-component field ${\bf E}$. 
Indeed, \CMRD{we can} decompose this field into a unit polarization vector and a scalar part: ${\bf E}({\bf r})={\bf e}({\bf r})\, E({\bf r})$, ${\bf e}^*\!\cdot {\bf e} =1$. 
This decomposition is not unique but can be fixed by choosing ${\rm Arg}({\bf e}\cdot {\bf e})=0$. Then, the gradient operator in Eq.~(\ref{total_phase}) acts on both the scalar field $E({\bf r})$ and the complex polarization vector ${\bf e}({\bf r})$, while in Eq.~(\ref{dynamical_phase}), involving the squared field $\Psi  = {\bf{E}} \cdot {\bf{E}}$ 
\KB{(where the \CMRD{scalar} product is calculated in a Cartesian basis\CMRD{)}}, the information about the polarization is {\it erased} because ${\rm Arg}\, \Psi({\bf r}) = 2\, {\rm Arg}\, E({\bf r})$. 
It is worth noticing that for either 2D or 3D fields, ${\bf E} = ({E_x,E_y})$ or ${\bf E} = ({E_x,E_y,E_z})$, the scalar ${\bf e}\cdot {\bf e}$ determines the {\it eccentricity} of the polarization ellipse \cite{BerryDennis2001}, while the condition ${\rm Arg}({\bf e}\cdot {\bf e})=0$ aligns the orthogonal ${\rm Re}({\bf e})$ and ${\rm Im}({\bf e})$ vectors with the principal axes of the polarization ellipse [see Section~(\ref{Sec-Coriolis}) below]. 
\KB{This criterion becomes ambiguous for the case of circular polarization \CMRD{(${\bf e}\cdot {\bf e}=0$)}, which is consistent with the fact that points/lines of circular polarization are polarization singularities \cite{Nye1983,Dennis2009}.}

The geometric phase is the phase associated with the evolution of the {\it polarization} ${\bf e}({\bf r})$  along the contour \cite{Shapere_book,Vinitskii1990,Bhandari1997,Ben-Aryeh2004,Malykin2004,Bliokh2009,Bliokh2015NP}. 
It is therefore natural to associate the difference between the total phase (\ref{total_phase}) and the dynamical phase (\ref{dynamical_phase}) with the {\it geometric phase} in an arbitrary \CMRD{multi-component} field:
\begin{equation}
\label{geometric_phase}
{\Phi _{\rm G}} = \Phi  - {\Phi _{\rm D}}
={\rm Im} \oint { \left[{\bf e}^* \! \cdot \left( \nabla  \right) {\bf e}\right]} \cdot d{\bf r}~.
\end{equation}
It is easy to see that this phase is invariant \CMRD{under} the gauge transformations ${\bf{E}}( {\bf{r}} ) \to {\bf{E}}( {\bf{r}}) \exp\! \left[ {i\chi ({\bf{r}})} \right]$, and it vanishes in uniformly-polarized fields with ${\bf e}({\bf r}) = {\bf const}$, where $\Phi = \Phi _{\rm D}$.
In \CMRD{the following,} we show that in the cases of 2D (paraxial) and 3D fields, the phase (\ref{geometric_phase}) is intimately related to the geometric and topological properties of the inhomogeneous polarization distribution. 
Moreover, for \KB{the cases considered previously}, the \CMRD{succinct} general Eqs.~(\ref{total_phase})--(\ref{geometric_phase}) yield the known expressions for the geometric phase (e.g., the Pancharatnam-Berry phase in paraxial fields). 
Notably, as the total phase $\Phi$ does not generically contain singularities, and the dynamical phase $\Phi _{\rm D}$ generically has singularities, {\it the geometric phase $\Phi _{\rm G}$ generically possesses singularities, i.e., undergoes $\pm\pi$ jumps when the contour of integration crosses phase singularities of the field $\Psi$}.

\KB{It is important to note that the dynamical phase (\ref{dynamical_phase}) in our consideration is quantized and, thus, has some {\it topological} (global) properties described by the integer $N_{\rm D}$. This is because we consider evolution along a {\it closed} contour in real space, 
\CMRD{for all phases, as discussed above}. 
In many earlier works, a cyclic evolution in the parameter space was assumed only for the geometric phase, while the dynamical phase was calculated for a unidirectional non-cyclic evolution in time or space. For the wave propagation along a non-cyclic contour (e.g., the $z$-axis), our definition (\ref{dynamical_phase}) coincides with the ``usual'' dynamical phase.}

\KB{Note also that} the space of the evolution ${\bf r}$ can be either ${\bf r}=(x,y)$, or ${\bf r}=(x,y,z)$, or any other parameter space, because the form $\nabla \cdot d{\bf r}$ used in the phase expressions (\ref{total_phase})--(\ref{geometric_phase}) is just the differential along the contour, and the resulting integrals are independent of the \CMRD{parametrization} of this contour. 
For simplicity, we use planar $(x,y)$-contours in all examples. 
Furthermore, \CMRD{we always use closed contour integrals in phase equations, which distinguish} between quantized and non-quantized phases; however, one can also calculate the \CMRD{total phase change} along an open contour connecting different ${\bf r}$-points. 
\CMRD{Counterclockwise} motion along circular spatial contours in the $(x,y)$-plane is \CMRD{always assumed}; the opposite motion flips the signs of all the phases.

\begin{figure}
\centering
\includegraphics[width=\linewidth]{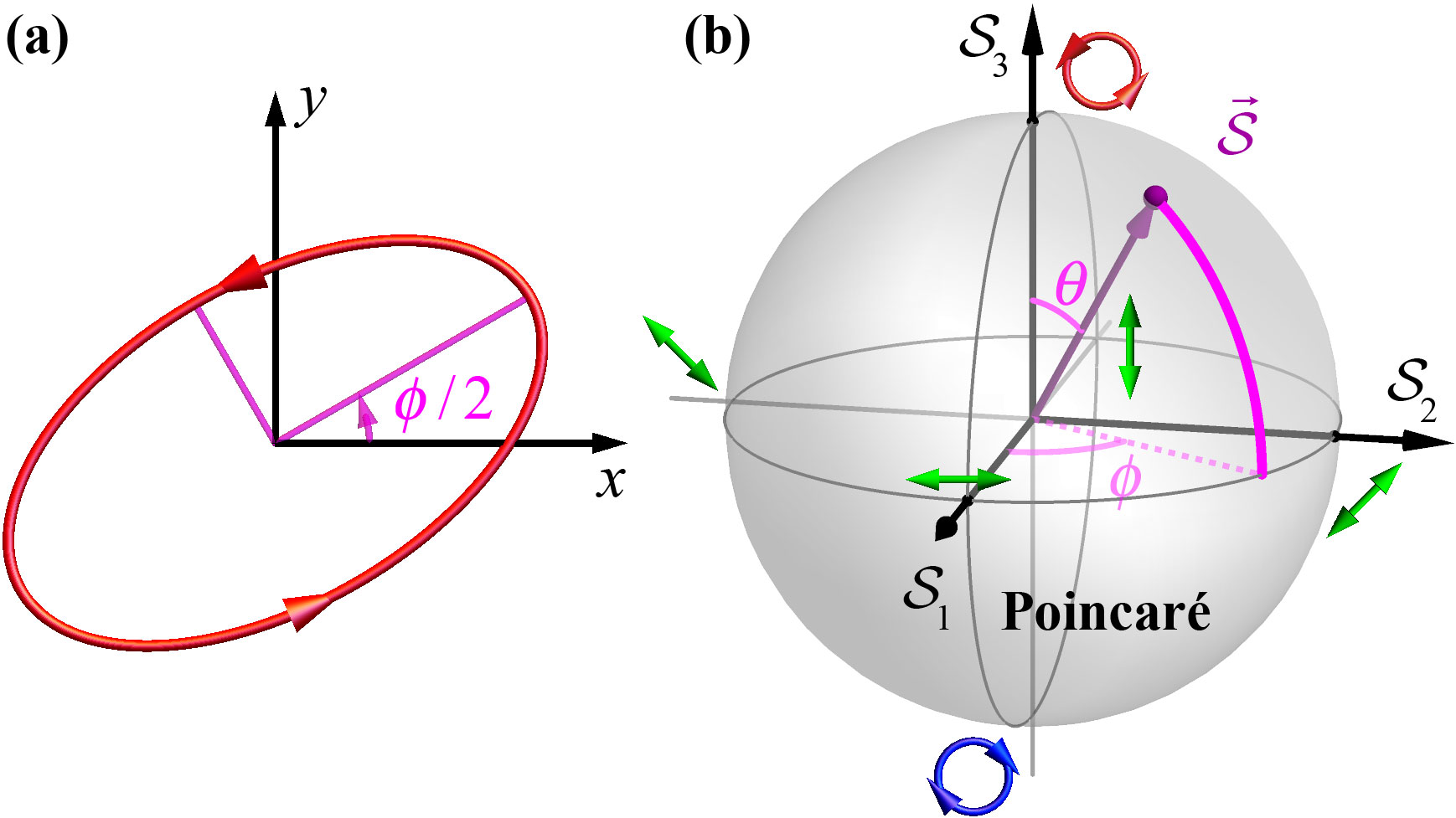}
\caption{(a) A generic form of the 2D polarization ellipse described by the complex field ${\bf E}({\bf r})$ at each point ${\bf r}$. (b) The Poincar\'{e} sphere (Stokes vector) representation of the polarization ellipse, Eqs.~(\ref{Stokes}) and (\ref{Poincare}) \cite{BornWolf,Azzam_book}. The polar and azimuthal angles on the Poincar\'{e} sphere, $\theta$ and $\phi$, describe the eccentricity and orientation of the polarization ellipse, respectively. \CKB{The magenta meridional line indicates the shortest geodesic line to the equator; its evolution determines the geometric phase in 2D fields, Eq.~(\ref{PB_phase}) and Fig.~\ref{phasejump}.}}
\label{Fig_Poincare}
\end{figure}

\subsection{\normalsize Paraxial 2D fields: the Pancharatnam-Berry phase and C-points}
\label{Pancharatnam}

We first apply the general formalism of Section~\ref{basic_idea} to the best-known case of a paraxial 2D field ${\bf{E}}\left( {\bf{r}} \right) = \left( {{E_x}\left( {\bf{r}} \right),{E_y}\left( {\bf{r}} \right)} \right)$, ${\bf{r}} = \left( {x,y} \right)$. 
This complex field describes the 2D {\it polarization ellipse}, which is traced by the temporal evolution of the real-valued electric field ${\rm Re} \left[ {{\bf{E}}\left( {\bf{r}} \right){e^{ - i\omega t}}} \right]$ at each point ${\bf r}$ \cite{BornWolf,Azzam_book}, Fig.~\ref{Fig_Poincare}(a). 
The geometric and topological properties of the \CMRD{smoothly-varying} polarization field of such ellipses \CMRD{are} the main subject of our consideration.
\KBII{Note that these} naturally appear in speckle patterns with random polarizations \cite{Freund2002,mrd2002,FSM2002,SDF2003,SDE2004,fodp2008}.

It is instructive to express the field ${\bf{E}}$ in the basis of circular polarizations ${{\bf{e}}^ \pm } = \left( {{\bf{\bar x}} \pm i\,{\bf{\bar y}}} \right)/\sqrt 2$, where ${\bf{\bar x}}$ and ${\bf{\bar y}}$ are the unit vectors for the corresponding Cartesian axes:
\begin{equation}
\label{2D_field}
\KB{{\bf{E}}^{\rm (circ)}} = \left( {\begin{array}{*{20}{c}}
{{E^ + }}\\
{{E^ - }}
\end{array}} \right) = \left( {\begin{array}{*{20}{c}}
{\left| {{E^ + }} \right|{e^{i{\alpha ^ + }}}}\\
{\left| {{E^ - }} \right|{e^{i{\alpha ^ - }}}}
\end{array}} \right).
\end{equation}
Here ${E^ \pm } = \left( {{E_x} \mp i\,{E_y}} \right)/\sqrt 2$, and $\alpha^\pm$ are the phases of the right-hand and left-hand field components \KB{(we use handedness for the view in the positive $z$\CMRD{-}direction)}. 

Substituting the field (\ref{2D_field}) into Eq.~(\ref{total_phase}), we obtain the total phase increment as:
\begin{equation}
\label{total_phase_2D}
\Phi  = \oint {\frac{{{{\left| {{E^ + }} \right|}^2}\nabla {\alpha ^ + } + {{\left| {{E^ - }} \right|}^2}\nabla {\alpha ^ - }}}{{{{\left| {{E^ + }} \right|}^2} + {{\left| {{E^ - }} \right|}^2}}} \cdot d{\bf{r}}}~.
\end{equation}
In turn, using the quadratic scalar field 
\begin{equation}
\label{quadratic_field_2D}
\Psi\!  ={E_x^2+E_y^2}
= 2{E^ + }{E^ - }\! = 
2\!\left| {{E^ + }} \right|\! \left| {{E^ - }} \right|\exp\! \left[{i \left( {{\alpha ^+}\! + {\alpha ^ - }} \right)} \right],
\end{equation}
the dynamical phase (\ref{dynamical_phase}) becomes:
\begin{equation}
\label{dynamical_phase_2D}
{\Phi _{\rm D}} = \frac{1}{2}\oint {\left( {\nabla {\alpha ^ + } + \nabla {\alpha ^ - }} \right) \cdot d{\bf{r}}}  = \pi N_{\rm D}~.
\end{equation}
The corresponding ``topological number'' is given by \CMRD{half the sum} of the topological charges of the right-hand and left-hand circular components of the field:
\begin{equation}
\label{dynamical_charge_2D}
\frac{N_{\rm D}}{2} = \frac{{{N^ + } + {N^ - }}}{2}~,
\end{equation}
where $\oint {\nabla {\alpha ^ \pm } \cdot d{\bf{r}}}  = 2\pi {N^ \pm }$. 

To show that the difference (\ref{geometric_phase}) between Eqs.~(\ref{total_phase_2D}) and (\ref{dynamical_phase_2D}) provides the known PB geometric phase, we introduce the {\it Poincar\'{e} sphere} representation of the 2D polarization \cite{BornWolf,Azzam_book}, Fig.~\ref{Fig_Poincare}. 
This sphere is determined by the normalized real-valued vector \CMRD{$\vec {\cal S}$ (denoted by an arrow as it lives in an abstract space, not real space)} of the normalized {\it Stokes parameters} \cite{BornWolf,Azzam_book}:
\begin{equation}
\label{Stokes}
\vec {\cal S} = \KB{\frac{{{{\bf{E}}^{\rm (circ)*}}\cdot \left( {\hat{\vec{\sigma}}} \right){\bf{E}}^{\rm (circ)}}}{{{{\left| {\bf{E}} \right|}^2}}}} = \left( {{{\cal S}_1},{{\cal S}_2},{{\cal S}_3}} \right).
\end{equation}
\KB{Here} 
${\hat{\vec{\sigma}}}=(\hat{\sigma}_1,\hat{\sigma}_2,\hat{\sigma}_3)$ is the vector of Pauli matrices, and $\vec {\cal S} \cdot \vec {\cal S} = 1$ (since we are considering fully polarized fields only). The north and south poles of the Poincar\'{e} sphere correspond to the right- and left-hand circular polarizations, whereas the equator represents all possible orientations of the linear polarization, as shown in  Fig.~\ref{Fig_Poincare}. Substituting the field (\ref{2D_field}) into Eq.~(\ref{Stokes}) yields the following spherical angles $(\theta,\phi)$ on the Poincar\'{e} sphere, Fig.~\ref{Fig_Poincare}(b):
\begin{eqnarray}
\label{Poincare}
&& \cos \theta = {{\cal S}_3} = \frac{{{{\left| {{E^ + }} \right|}^2} - {{\left| {{E^ - }} \right|}^2}}}{{{{\left| {\bf{E}} \right|}^2}}}~,\nonumber \\
&& \phi = {\mathop{\rm Arctan}\nolimits} \frac{{{{\cal S}_2}}}{{{{\cal S}_1}}} = {\mathop{\rm Arg}\nolimits} \left( {{E^{ + *}}{E^ - }} \right) = {\alpha ^ - } - {\alpha ^ + }.
\end{eqnarray}
%

Now, using Eqs.~(\ref{total_phase_2D}), (\ref{dynamical_phase_2D}), and (\ref{Poincare}), we can write the geometric phase (\ref{geometric_phase}) in a 2D field as
\begin{eqnarray}
\label{PB_phase}
{\Phi _{\rm G}} & = & \frac{1}{2}\oint {\frac{{{{\left| {{E^ + }} \right|}^2} - {{\left| {{E^ - }} \right|}^2}}}{{{{\left| {{E^ + }} \right|}^2} + {{\left| {{E^ - }} \right|}^2}}}} \,\left( {\nabla {\alpha ^ + } - \nabla {\alpha ^ - }} \right) \cdot d{\bf{r}} \nonumber\\
& = & -\, \frac{1}{2}\oint {\cos \theta } \,d\phi = \frac{1}{2}\,\Sigma_{\rm equat}~.
\end{eqnarray}
\CKB{The second equality in this equation means that the geometric phase, originally defined via an integration in real space, can be calculated as an integral along the contour traced by the representation point 
\CMRD{on the Poincar\'{e} sphere corresponding to the sequence of polarization ellipses on the ${\bf r}$-contour}. 
In turn, the last equality in Eq.~(\ref{PB_phase}) indicates that the geometric phase is numerically equal to half of the spherical area $\Sigma_{\rm equat}$ swept by the shortest geodesic line connecting the point on the contour and the equator of the Poincar\'e sphere (see Fig.~\ref{Fig_Poincare}(b)); for the motion in the positive (negative) $\phi$-direction the northern-hemisphere area is counted as negative (positive), while the southern-hemisphere area is counted with the opposite signs. }

\begin{figure}[t!]
\centering
\includegraphics[width=\linewidth]{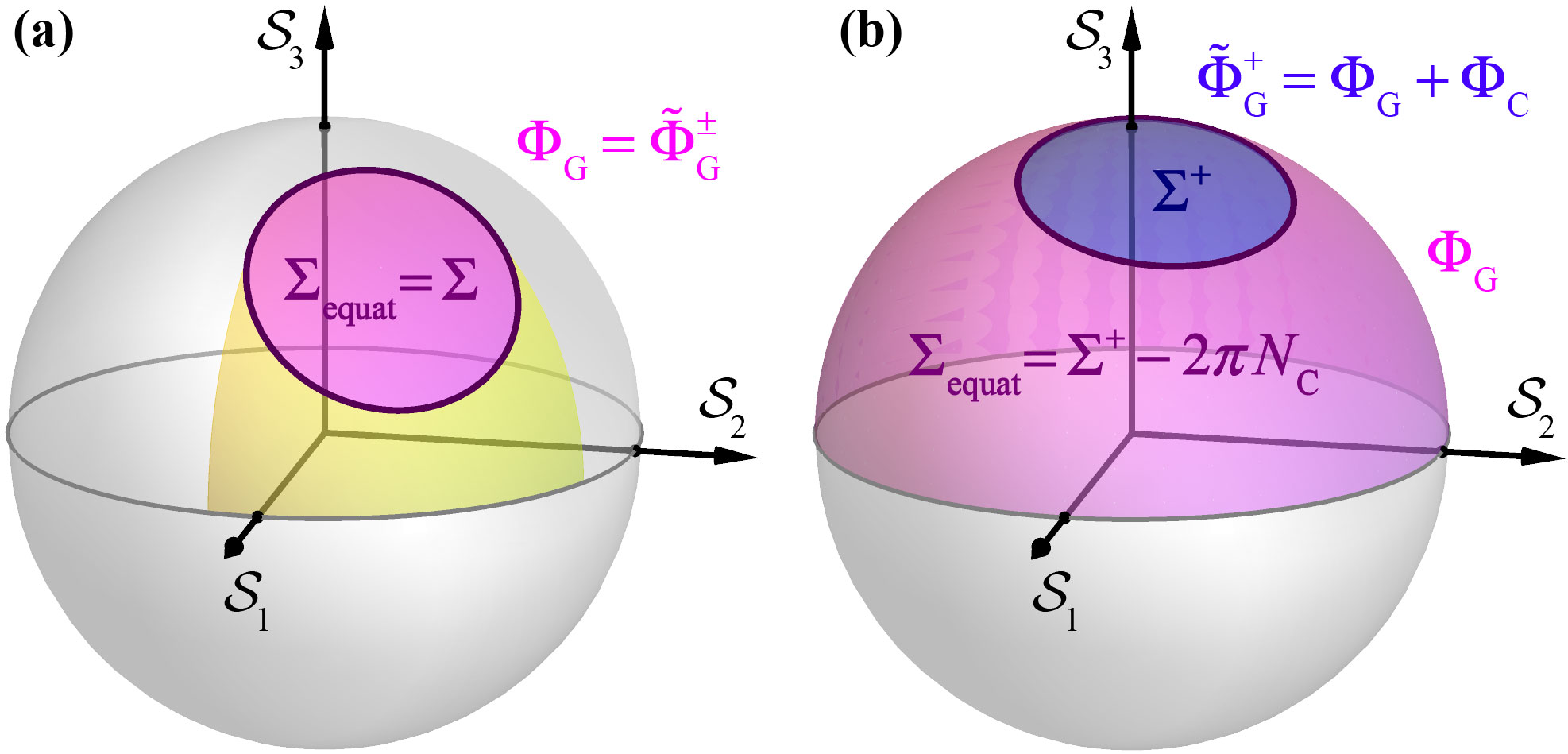}
\caption{\CKB{Geometric and topological properties of the PB phase on the Poincar\'e sphere, Eqs.~(\ref{PB_phase}), (\ref{area}), and (\ref{PB_phasePrime}). The geometric phase $\Phi_{\rm G}$ is equal to half of the oriented area swept by the shortest geodesic line to the equator [shown in Fig.~\ref{Fig_Poincare}(b)], $\Sigma_{\rm equat}$, as described in the text. (a) For a contour that does not enclose the $\mathcal{S}_3$ axis of the Poincar\'{e} sphere (i.e., the C-points of circular polarizations in real space), the PB phase is uniquely defined and corresponds to half the oriented area (solid angle) $\Sigma$ enclosed by the contour on the Poincar\'{e} sphere and not including the Poincar\'{e}-sphere poles (shown in magenta). The area between the contour and equator (shown in yellow) is swept twice in opposite directions and hence does not contribute. 
(b) For a contour \CMRD{enclosing} the $\mathcal{S}_3$ axis, the oriented area $\Sigma_{\rm equat}$ swept by the geodesic to the equator (shown in magenta) differs from the area $\Sigma^{\pm}$ enclosed by the contour (shown in blue, ``$+$'' and ``$-$'' indicating the areas including the north and south poles, respectively) by $2\pi N_{\rm C}$. Here $N_{\rm C}$ is the net topological index of the C-points enclosed by the contour in real space or, equivalently, the winding number of the contour around the $\mathcal{S}_3$ axis on the Poincar\'e sphere. These two areas correspond to the fundamental and modified \CMRD{definitions} of the PB geometric phase, Eqs.~(\ref{PB_phase}), (\ref{area}), and (\ref{PB_phasePrime}).}}
\label{phasejump}
\end{figure}

\CKB{The Poincar\'{e}-sphere form of the geometric phase (\ref{PB_phase}) provides the well-known  {\it Pancharatnam-Berry (PB) phase} \cite{Vinitskii1990,Bhandari1997,Ben-Aryeh2004,Berry1987,Bhandary1988,Simon1988}, which has many important applications in modern optics, such as ``metasurfaces'' and ``q-plates'' providing efficient polarization (spin) -dependent shaping of optical beams
\cite{Hasman2005,Marrucci2011,Bliokh2015NP,Capasso2014,Chen2016,Xiao2017,
Maguid2016,Maguid2017,Alonso2019,Rubano2019}.
The PB phase is usually associated with the oriented area (solid angle) enclosed by the contour on the Poincar\'e sphere, but there is a subtle yet important issue with its definition (\ref{PB_phase}). 
When the Poincar\'e-sphere contour does not enclose the poles, i.e., does not wind around the $\mathcal{S}_3$ axis, the area swept by the geodesic to the equator becomes equal to the oriented area (or equivalently the solid angle) enclosed by the contour: $\Sigma_{\rm equat} = \Sigma$, Fig.~\ref{phasejump}(a).  
However, each time that the contour \CMRD{encloses} the $\mathcal{S}_3$ axis of the Poincar\'e sphere, 
the area $\Sigma_{\rm equat}$ swept by the geodesic differs from the area $\Sigma^+$ enclosed by the contour and including the north pole by $\mp 2\pi$ (where the sign is determined by the direction of winding around the $\mathcal{S}_3$ axis), Fig.~\ref{phasejump}(b). Therefore, {\it the geometric phase $\Phi_{\rm G}$ experiences a $\pm\pi$ jump when the contour crosses a pole on the Poincar\'e sphere}. This singular behaviour is intimately related to the singularities and {\it topological} properties of spatial polarization distributions.} 

\begin{figure}[t!]
\centering
\includegraphics[width=\linewidth]{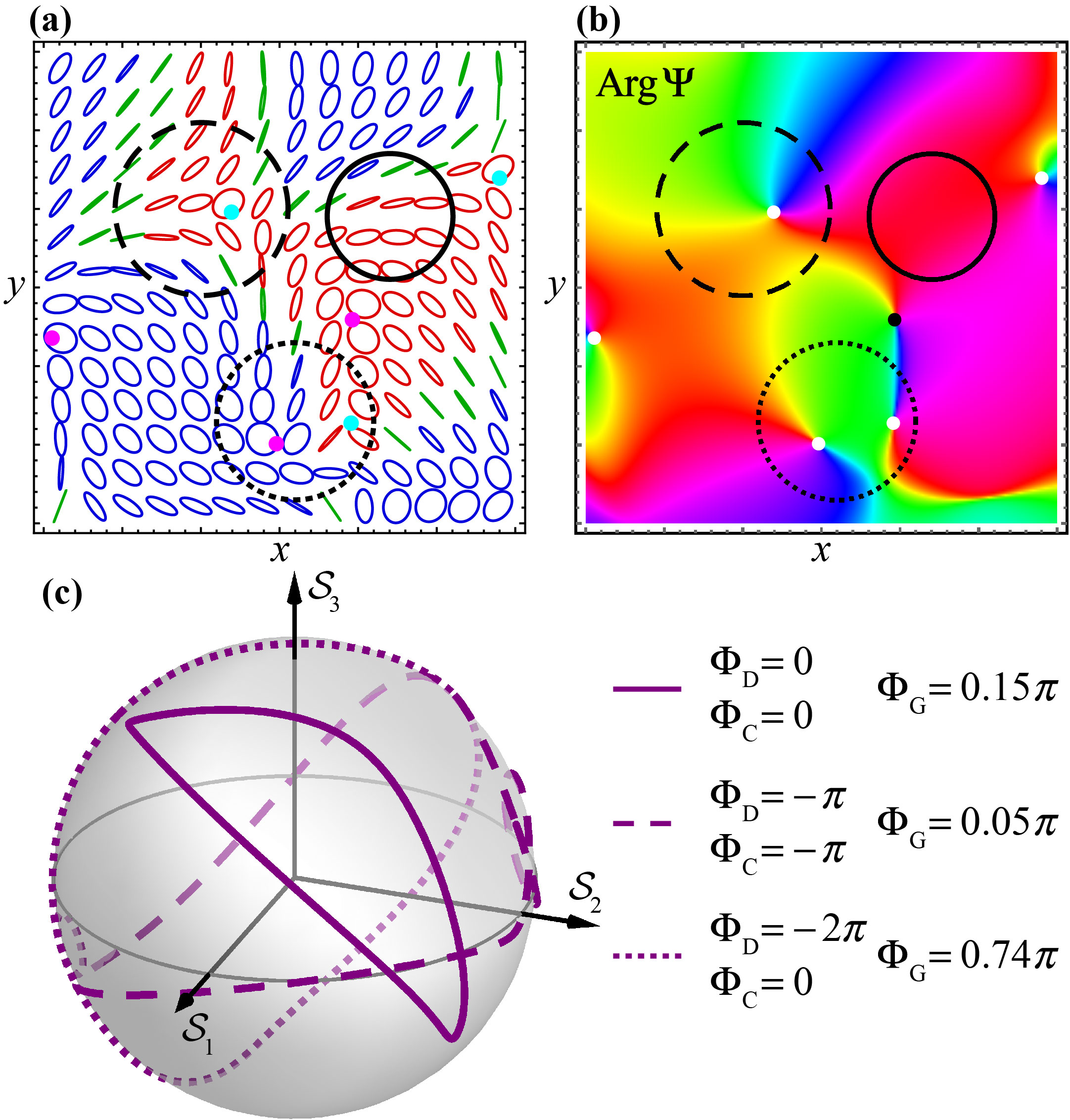}
\caption{An example of a random 2D polarized field ${\bf E}({\bf r})$ with polarization singularities and geometric (Pancharatnam-Berry) / dynamical / C-point phases calculated along circular contours in this field. (a) Spatial distribution of the normalized polarization ellipses. Red, blue, and green colours correspond to the right-handed ($\mathcal{S}_3>0$), left-handed ($\mathcal{S}_3<0$), and near-linear ($\mathcal{S}_3 \simeq 0$) polarizations. Polarization singularities (C-points of purely circular polarization) with positive and negative topological indices $N_{\rm C}/2 = \pm 1/2$ (indicating $\pm \pi$ rotations of the polarization-ellipse orientation when going counterclockwise around the singularity) are marked by the magenta and blue dots, respectively \cite{Nye1983,Dennis2009}. (b) The phase distribution of the quadratic scalar field $\Psi = {\bf E} \cdot {\bf E}$ with its phase singularities marked as in Fig.~\ref{Fig_scalar}. (c) Evolution of the polarization ellipses along the three contours shown in (a,b) represented on the Poincar\'{e} sphere. The dynamical,  geometric, and C-point phases, Eqs.~(\ref{dynamical_phase_2D}), (\ref{PB_phase}), (\ref{C_phase_2D}), and (\ref{area}), calculated for these contours are shown.}
\label{Fig_random_2D}
\end{figure}

\CKB{Figures~\ref{Fig_random_2D}(a,b) show} a generic example of an inhomogeneous polarization distribution in a random 2D field ${\bf{E}}({\bf{r}})$ and the phase distribution in the corresponding scalar field $\Psi  = {\bf{E}} \cdot {\bf{E}}$. The phase singularities in the scalar field $\Psi$ correspond to the points of purely-circular polarizations in the field ${\bf{E}}$. These are {\it polarization singularities} (the orientation of the ellipse is undetermined) called ``{\it C-points}'' \cite{Dennis2009,Nye1983}. Importantly, in the generic case the directions of the principal axes of the polarization ellipse in the vicinity of a C-point undergo a $\pm \pi$ rotation (half-turn) when going counterclockwise along a contour enclosing the singularity. This determines a half-integer {\it topological index} $N_{\rm C}/2 = \pm 1/2$ of the C-point \cite{Dennis2009,Nye1983}. Akin to the phase singularities, higher-order C-points are possible in degenerate cases (see an example in Fig.~\ref{vortex} below). Notably, the topological indices of C-points [indicated by the magenta and cyan dots in Fig.~\ref{Fig_random_2D}(a)] and the topological charges $N_{\rm D}=\pm 1$ of phase singularities of the field $\Psi$ [indicated by the black and white dots in Fig.~\ref{Fig_random_2D}(b)] are independent of each other. Thus, one can characterize each polarization \CMRD{singularity} of a 2D field by {\it two integer topological numbers} $(N_{\rm D},N_{\rm C})$. 

\CMRD{The} first topological number $N_{\rm D}$ \CMRD{obviously} corresponds to the dynamical phase (\ref{dynamical_phase_2D}) for the contour enclosing the singularity. 
One can show that the number $N_{\rm D}/2$ counts {\it the number of turns of the direction of the instantaneous field vector ${\rm Re} [{\bf{E}}({\bf{r}})]$ with respect to the major axis of the polarization ellipse} when going along the contour. The second number $N_{\rm C}/2$ counts {\it the number of turns of the direction of the major axis of the polarization ellipse} itself. Since the orientation of the major axis of the polarization ellipse with respect to the Cartesian $(x,y)$-axes is given by the half-azimuthal angle $\phi/2$ on the Poincar\'{e} sphere (see Fig.~\ref{Fig_Poincare}), one can introduce the quantized phase corresponding to the topological index $N_{\rm C}/2$ for the C-points enclosed by the contour:
\begin{equation}
\label{C_phase_2D}
{\Phi _{\rm C}} = \frac{1}{2}\oint {d\phi }  = \frac{1}{2}\oint {\left( {\nabla {\alpha ^ - } - \nabla {\alpha ^ + }} \right)} \, \cdot d{\bf{r}}=\pi N_{\rm C}~.
\end{equation}
\KBII{Note that the phase angle $\phi$ is the argument of the complex scalar $\mathcal{S}_1 + i\, \mathcal{S}_2 = 2\,E^{+*}E^-$, whose zeros, the ``Stokes vortices'' are again the C-points \cite{mrd2002,FSM2002,FMSAM2002}.The difference with the phase $\Phi_{\rm D}$ coming from $\Psi = 2 E^+ E^-$ is related to the conjugation of $E^+$.}

\CMRD{In analogy to} Eq.~(\ref{dynamical_charge_2D}), one can see that the topological index $N_{\rm C}/2$ is given by \CMRD{half the difference} of the topological charges of the left-handed and right-handed circular components of the field:
\begin{equation}
\label{C_charge_2D}
\frac{{{N_{\rm C}}}}{2} = \frac{{{N^ - } - {N^ + }}}{2}~.
\end{equation}
\CKB{Clearly, $N_{\rm D}$ and $N_{\rm C}$ have the same parity, so that
\begin{equation}
\label{parity}
N_{\rm C}\, {\rm mod}\, 2 = N_{\rm D}\, {\rm mod}\, 2, ~~
\Phi_{\rm C}\, {\rm mod}\, 2\pi = \Phi_{\rm D}\, {\rm mod}\, 2\pi.
\end{equation}
}
Figure~\ref{Fig_random_2D} shows examples of spatial contours enclosing zero, one, or two C-points with different topological numbers $(N_{\rm D},N_{\rm C})$ and the corresponding quantized phases $\Phi_{\rm D}$ and $\Phi_{\rm C}$, Eqs.~(\ref{dynamical_phase_2D}) and (\ref{C_phase_2D}).

\begin{figure*}
\centering
\includegraphics[width=0.8\linewidth]{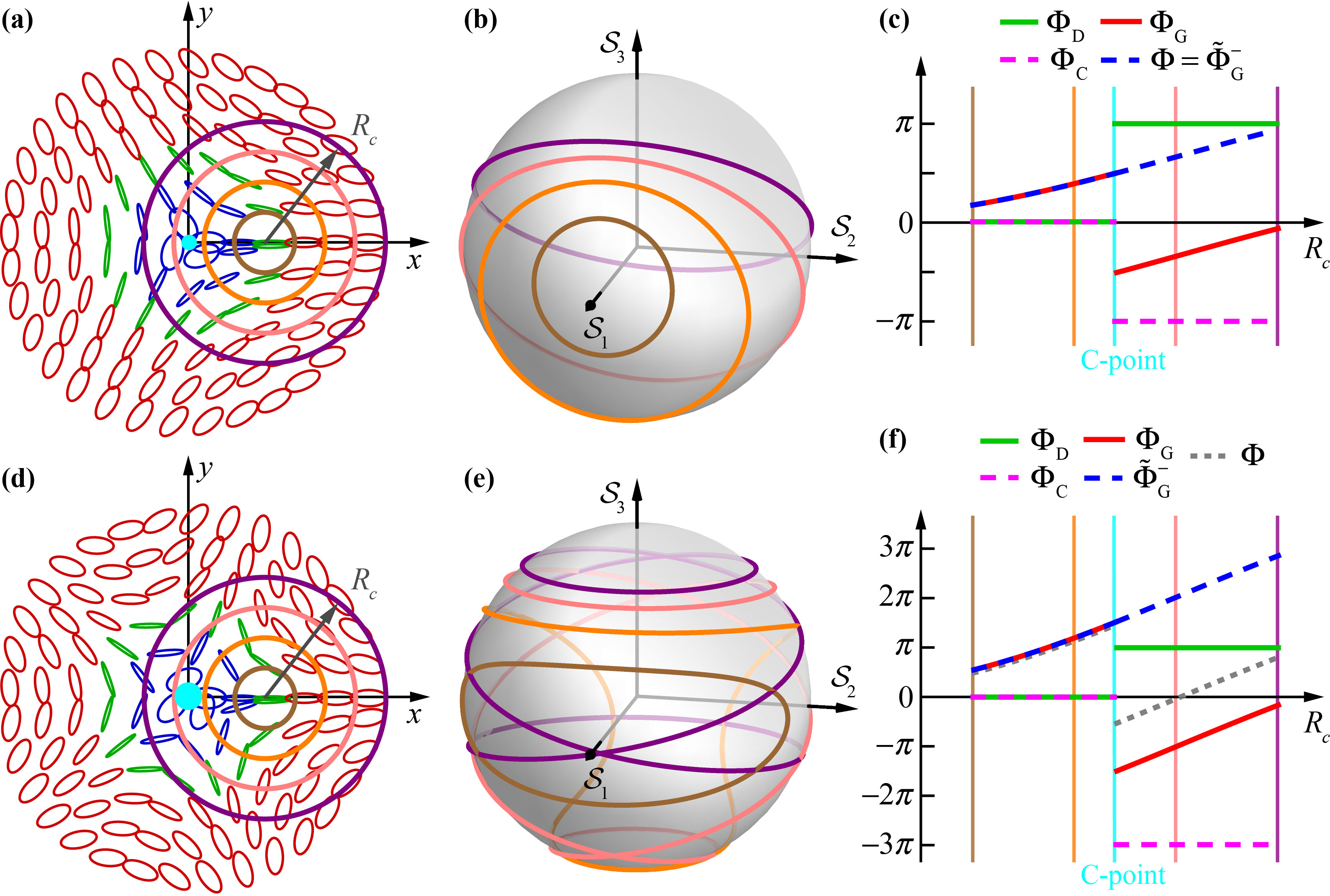}
\caption{Various phases in 2D ``polarization vortices'', i.e., polarization distributions with discrete azimuthal symmetries and a C-point at the center. C-points (polarization vortices) with topological charges $N_{\rm C}/2=-1/2$ and $-3/2$ are shown in the panels (a--c) and (d--f), respectively. (a,d) Spatial distributions of the normalized polarization ellipses with the color convention analogous to Fig.~\ref{Fig_random_2D}(a), C-points (cyan dots), and a family of off-axis circular contours with the varying radius $R_c$. (b,e) The representations of the spatial contours from (a,c) on the Poincar\'e sphere. (c,f) The dynamical, C-point, total, and geometric PB (according to different definitions) phases, numerically calculated from Eqs.~(\ref{total_phase_2D}), (\ref{dynamical_phase_2D}), (\ref{PB_phase}), (\ref{C_phase_2D}), (\ref{area}), and (\ref{PB_phasePrime}). See explanations of their behaviour in the text.}
\label{vortex}
\end{figure*}

Coming back to the PB geometric phase (\ref{PB_phase}), Fig.~\ref{Fig_random_2D}(c) shows the Poincar\'e-sphere contours representing the circular spatial contours shown in Figs.~\ref{Fig_random_2D}(a,b) and the corresponding numerically-calculated geometric phases $\Phi_{\rm G}$. \CKB{Note that the topological number $N_{\rm C}$, counting the C-points enclosed by the contour in real space, equals {\it the number of times the Poincar\'e-sphere contour winds around the vertical $\mathcal{S}_3$-axis}. 
From this relation and properties of the PB phase (\ref{PB_phase}) discussed above and shown in Fig.~\ref{phasejump}, the geometric phase can be written as} 
\begin{equation}
\label{area}
\Phi_{\rm G} = \frac{1}{2}\Sigma^{\pm} \mp \pi N_{\rm C}, 
\end{equation}
where $\Sigma^\pm$ denote the oriented areas (or, equivalently, the solid angles) on the Poincar\'e sphere enclosed by the contour. 
Here, the $\pm$ index distinguishes the areas {\it above} and {\it below} the contours \CMRD{enclosing} the $\mathcal{S}_3$ axis, i.e., including the north and south poles, respectively; 
for contours not enclosing the $\mathcal{S}_3$ axis ($N_{\rm C}=0$), $\Sigma^+ = \Sigma^- = \Sigma$ is the area not including the poles, see Fig.~\ref{phasejump}.

\CKB{Thus, the geometric phase $\Phi_{\rm G}$ experiences $\pm \pi$ jumps when the real-space contour crosses a generic C-point.} This \CMRD{agrees} with the general properties mentioned in Section~\ref{basic_idea}, and only the total phase $\Phi$ evolves continuously.
Due to this, the PB phase in paraxial fields is often defined via the following modified expression (typically, with the ``+'' sign) \cite{Vinitskii1990,Bhandari1997,Ben-Aryeh2004,Berry1987,Bhandary1988,Simon1988}:
\begin{equation}
\label{PB_phasePrime}
{\tilde{\Phi}_{\rm G}^\pm}  =  {\Phi _{\rm G}} \pm {\Phi _{\rm C}} = \frac{1}{2}\oint (\pm 1-{\cos \theta })\,{d\phi } = \frac{1}{2}\Sigma^{\pm}. 
\end{equation}
This definition, determined only by the corresponding oriented area enclosed by the contour, is more convenient in some practical situations. For example, the ${\tilde{\Phi}_{\rm G}^+}$ phase does not experience jumps when the Poincar\'e-sphere contour crosses the north pole, and experience a $2\pi$ jump when it crosses the south pole. Therefore, in all situations where only the phase modulo $2\pi$ is observable (e.g., in numerous interference experiments), this \CMRD{definition} becomes effectively invariant with respect to global SU(2) transformations of the field (or, equivalently SO(3) rotations of the contour on the Poincar\'e sphere), resulting, e.g., from making the complete paraxial field pass through a uniform transparent waveplate retarder. \CKB{It also follows from Eq.~(\ref{parity}) that}
\[
\tilde{\Phi}_{\rm G}^+ \, {\rm mod}\, 2\pi = \tilde{\Phi}_{\rm G}^- \,{\rm mod}\, 2\pi = \Phi \,{\rm mod}\, 2\pi~.
\] 
\KBII{This is the reason why the total phase $\Phi$ is sometimes confused with the geometric PB phase \cite{Nye1991,BerryDennis2001}.} 

However, it is important to emphasize that the geometric phase $\Phi_{\rm G}$, Eqs.~(\ref{PB_phase}) and  (\ref{area}), is more fundamental and consistent with the universal expression (\ref{geometric_phase}) for an arbitrary multicomponent field. In particular, we will see in Section~\ref{Sec-General3D} that the natural definitions of the geometric phase for 3D fields is consistent with $\Phi_{\rm G}$ rather than with ${\tilde{\Phi}_{\rm G}^\pm}$, and the discontinuous nature of $\Phi_{\rm G}$ near polarization singularities will be related to the polarization M\"obius-strip structures.
 
We conclude this section with an explicit example of an inhomogeneous 2D polarized field, which plays an important role in modern optics. This is a transverse $(E_x,E_y)$ field of a $z$-propagating cylindrical paraxial beam with a C-point (\CMRD{of first or higher-order}) in the center. The $(x,y)$ polarization distribution forms a ``{\it polarization vortex}'' around the field center (see Fig.~\ref{vortex}); such polarization distributions appear in a variety of optical systems \cite{Freund2002,Niv2006,Maurer2007,Wang2007,Bliokh2008OE,Zhan2009,Alonso2010,
Milione2011,Galvez2012,Cardano2013,Otte2016,Zhang2018,Doeleman2018_1}. 
The polarization-vortex beam field represents a superposition of two vortex beams \cite{Allen_book,Andrews_book,Molina2007,Franke2008,Bliokh2015PR} with opposite circular polarizations and different vortex charges. In terms of the field (\ref{2D_field}), this means $\alpha^{\pm} = \ell^{\pm} \varphi$, $|E^{\pm}| \propto \rho^{|\ell^{\pm}|}$ (we consider the field in the vicinity of the beam axis), where $\ell^{\pm}$ are the integer vortex charges, and $(\rho,\varphi)$ are the polar coordinates in the $(x,y)$ plane. For circular contours enclosing the C-point in the beam center, we have $N^{\pm} = \ell^{\pm}$, and the difference between the dynamical-phase and C-point properties (\ref{dynamical_phase_2D}), (\ref{dynamical_charge_2D}) and (\ref{C_phase_2D}), (\ref{C_charge_2D}) can be easily appreciated: $N_{\rm D} = \ell^+ + \ell^-$ and $N_{\rm C} = \ell^- - \ell^+$. 
Thus, the topological charge of the C-point (polarization vortex) equals $(\ell^- - \ell^+)/2$, and the cases with the charges $-1/2$ ($\ell^+=1$, $\ell^-=0$) and $-3/2$ ($\ell^+=2$, $\ell^-=-1$), i.e., generic and degenerate, are shown in Figs.~\ref{vortex}(a--c) and (d--f), respectively. 

To trace the phase changes when the contour crosses the C-point, we now consider a set of circular contours of different radii centered some distance away from the beam center, as shown in Fig.~\ref{vortex}(a,d). The Poincar\'e-sphere representations of these contours are shown in Figs.~\ref{vortex}(b,e), while Figs.~\ref{vortex}(c,f) display the numerically-calculated dynamical, C-point, total, and geometric PB phases as functions of the contour radius $R_c$. One can clearly see that $\Phi_{\rm D}=\Phi_{\rm C}=0$ when the contour does not enclose the C-point and different quantized values of $\Phi_{\rm D}$ and $\Phi_{\rm C}$ after crossing the C-point. Furthermore, all \CMRD{definitions} of the PB geometric phase (\ref{PB_phase}), (\ref{area}), and (\ref{PB_phasePrime}) coincide with each other and with the total phase (\ref{total_phase_2D}), $\Phi_{\rm G}={\tilde{\Phi}}_{\rm G}^\pm =\Phi$, and grow with the contour radius before crossing the C-point. Upon the contour crossing the C-point, the fundamental geometric phase $\Phi_{\rm G}$ experiences a $\pi N_{\rm C}$ jump in agreement with Eq.~(\ref{area}). In turn, the modified PB phase ${\tilde{\Phi}}_{\rm G}^-$ evolves continuously, because in these examples the C-points have left-hand circular polarizations corresponding to the south pole of the Poincar\'e sphere. Note also that the total phase $\Phi$ evolves continuously through the generic C-point in Fig.~\ref{vortex}(a--c), but experiences a $-2\pi$ jump in the degenerate C-point in Fig.~\ref{vortex}(d--f), because the field components $E^+$ and $E^-$ vanish simultaneously at this point.

\section{\normalsize 3D polarized fields}
\label{Sec-General3D}
\subsection{\normalsize Geometric phase as the Coriolis effect. Spin-redirection phase}
\label{Sec-Coriolis}

\begin{figure}
\centering
\includegraphics[width=0.6\linewidth]{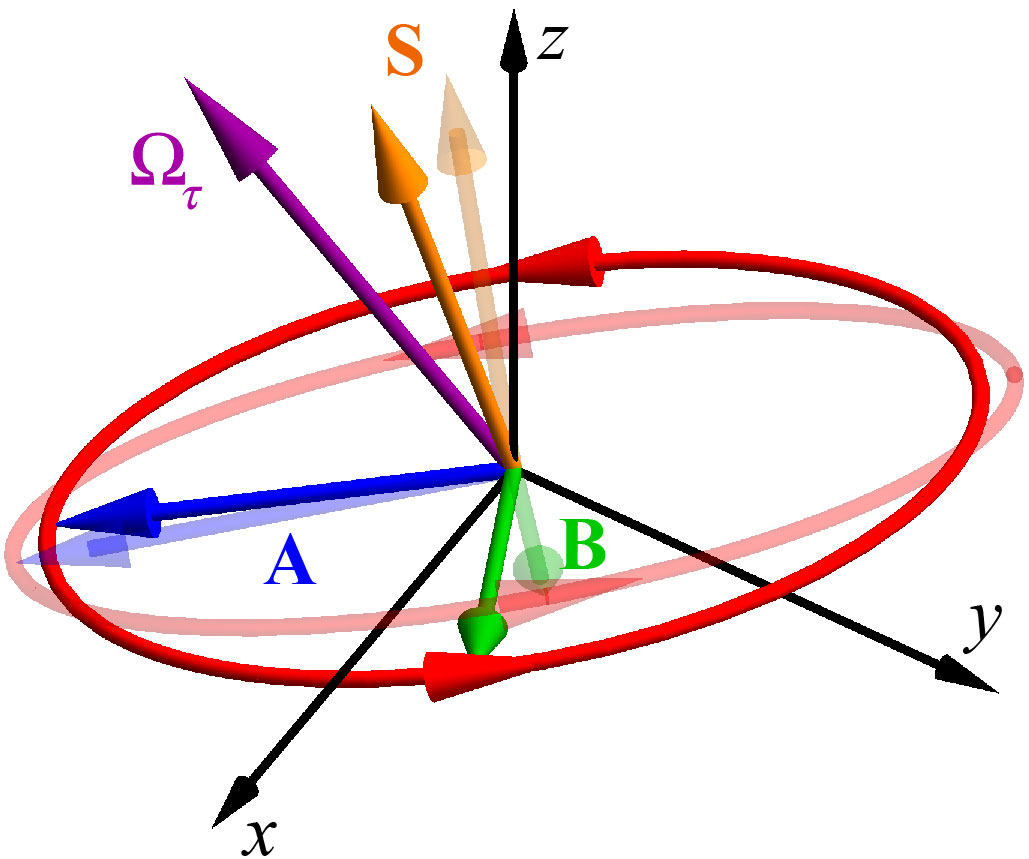}
\caption{Schematics of the 3D polarization ellipse (\ref{AB}) with semiaxes vectors ${\bf A}$ and ${\bf B}$, as well as the normal ``spin'' vector (\ref{spin}) ${\bf S}$. The two polarization ellipses shown here are taken from the neighbouring points on the contour, which can be marked by the parameter $\tau=\tau_0$ (opaque) and $\tau=\tau_0 + \delta\tau$ (semitransparent). The corresponding evolution of the normalized triad ($\bar{\bf a}$,$\bar{\bf b}$,$\bar{\bf s}$) can be presented as a rotation with angular velocity $\bm{\Omega}_\tau$, so that the vector angle of rotation between the two neighboring points is $\delta\bm{\alpha}=\bm{\Omega}_\tau\, \delta\tau$. The geometric phase acquired during this evolution is determined by the Coriolis spin-rotation coupling (\ref{Coriolis}): $\delta\Phi_{\rm G} = - \bm{S}\cdot \bm{\Omega}_\tau\, \delta\tau$.}
\label{Fig_Coriolis}
\end{figure}

We are now in a position to consider generic 3D fields ${\bf{E}}\left( {\bf{r}} \right)=\left( {{E_x}(\br),{E_y}(\br),{E_z}(\br)} \right)$, ubiquitous in modern optics. 
\CMRD{As we cannot use Stokes parameters in 3D,} it
is convenient to use the following representation of the complex vector field \CMRD{\cite{BornWolf,Nye1987,Nye1991,mrd2002,BerryDennis2001}}:
\begin{equation}
\label{AB}
{\bf{E}} = |{\bf{E}}| \left( {{\bf{A}} + i\,{\bf{B}}} \right){e^{i\epsilon }} = |{\bf{E}}|\, {\bf{e}}\,{e^{i\epsilon }}\,.
\end{equation}
where ${\bf A}$, ${\bf B}$, and $\epsilon$ are real fields, such that ${\bf A}\cdot {\bf B} =0$,  $|{\bf A}| \geq |{\bf B}|$, and (unlike previous publications \cite{BornWolf,Nye1987,Nye1991,BerryDennis2001}) we use normalized quantities ${\bf e}^*\!\cdot {\bf e} = |{\bf A}|^2+|{\bf B}|^2 =1$. The real vectors ${\bf A}$ and ${\bf B}$ correspond to the {\it major and minor semi-axes of the normalized polarization ellipse} (whose definition is unique to within a global sign, except at C-points), Fig.~\ref{Fig_Coriolis}. Equation (\ref{AB}) provides a natural decomposition of the field ${\bf E}$ into its real amplitude $|{\bf{E}}|\equiv \sqrt{{\bf E}^*\!\cdot {\bf E}}$, the complex polarization field ${\bf e}={{\bf{A}} + i\,{\bf{B}}}$ (which is in agreement with the general polarization definition of Section~\ref{basic_idea}), ${\rm Arg} ({\bf e} \cdot {\bf e}) = 0$), and the common polarization-independent phase $\epsilon$, \CMRD{known as the ``phase of the vibration'' \cite{BornWolf,Nye1987,Nye1991,mrd2002}}. 

Substituting representation (\ref{AB}) into the general Eqs.~(\ref{total_phase}) and (\ref{dynamical_phase}), we obtain the total and dynamical phases:
\begin{eqnarray}
\label{phase_3DT}
\Phi & = & \oint {\left[ {\nabla \epsilon  + 2\,{{{\bf{A}} \cdot \left( \nabla  \right){\bf{B}}}}} \right] \cdot d{\bf{r}}}\, ,\\
\label{phase_3DD}
{\Phi _{\rm D}} & = & \oint {\nabla \epsilon  \cdot d{\bf{r}}} = \pi N_{\rm D}~,
\end{eqnarray}
where we used ${\bf{A}} \cdot \left( \nabla  \right){\bf{B}} = - {\bf{B}} \cdot \left( \nabla  \right){\bf{A}}$. Equation (\ref{phase_3DD}) elucidates the physical meaning of the dynamical phase: it can be regarded as the phase in the local coordinate frame attached to the polarization ellipse's axes ${\bf A}$ and ${\bf B}$. In other words, as mentioned in Section~\ref{Pancharatnam}, it corresponds to the number of turns of the instantaneous electric-field vector ${\rm Re} ({\bf{E}}) = |{\bf{E}}|\left({\bf{A}}\cos \epsilon  - {\bf{B}}\sin \epsilon \right)$ with respect to these coordinates. 
Equations~(\ref{phase_3DT}) and (\ref{phase_3DD}) immediately yield the geometric phase in the form (\ref{geometric_phase}):
\begin{equation}
\label{geometric_phase_3D}
{\Phi _{\rm G}} = 2 \oint {{{ \left[{\bf{A}} \cdot \left( \nabla  \right){\bf{B}}\right] }} \cdot d{\bf{r}}}
= {\rm Im} \oint {\left[ {\bf e}^*\! \cdot \left( \nabla  \right){\bf e} \right]} \cdot d{\bf r}.
\end{equation}
To show that Eq.~(\ref{geometric_phase_3D}) coincides with prior definitions of the geometric phase in 3D fields, we represent it in the form of the {\it Coriolis (or rotational-Doppler) coupling between the intrinsic angular momentum carried by the wave and the coordinate-frame rotations} \cite{Bliokh2008,Garetz1981,Mashhoon1988,Lipson1990,Bretenaker1990, Courtial1998,Bliokh2009}. 

Recall that a wave carrying normalized intrinsic angular momentum ${\bf S}$ and observed in a coordinate frame, which rotates  with angular velocity ${\bm \Omega}_\tau$ (\CMRD{parametrized by} $\tau$, not necessarily time), acquires the following geometric-Coriolis-Doppler phase shift \cite{Bliokh2008,Garetz1981,Mashhoon1988,Lipson1990,Bretenaker1990, Courtial1998,Bliokh2009}:
\begin{equation}
\label{Coriolis}
{\Phi _{\rm G}} =  - \int {{\bf{S}} \cdot {{\bm\Omega}_\tau }} \,d\tau~.
\end{equation}
The best-known example of this phase is provided by the rotation of the Foucault pendulum on Earth's rotating surface \cite{Berry1988,Opat1991}. 
For the \CMRD{wavefield} under consideration, the intrinsic angular momentum is the normalized ``spin density'' determined by the ellipticity and the normal to the polarization ellipse \CMRD{\cite{Bliokh2015PR,BerryDennis2001}} (not to be confused with the Stokes vector $\vec{\cal S}\,$!), Fig.~\ref{Fig_Coriolis}: 
\begin{equation}
\label{spin}
{\bf{S}} = \frac{{{\rm Im} \left( {{{\bf{E}}^*}\! \times {\bf{E}}} \right)}}{{{{\left| {\bf{E}} \right|}^2}}}
= {{{\rm Im} \left( {{{\bf{e}}^*}\! \times {\bf{e}}} \right)}} 
= 2\,{\bf{A}} \times {\bf{B}}~.
\end{equation}
The triad $({\bf A},{\bf B},{\bf S})$ determines {\it the local Cartesian frame attached to the polarization ellipse}, whose unit vectors can be defined as ${\bf{\bar a}} = {\bf{A}}/\left| {\bf{A}} \right|$, ${\bf{\bar b}} = {\bf{B}}/\left| {\bf{B}} \right|$, and  ${\bf{\bar s}} = {\bf{S}}/\left| {\bf{S}} \right|$
\CMRD{\cite{Nye1987,BerryDennis2001}}.
The rotation of the vectors $({\bf A},{\bf B})$ with respect to the normal direction ${\bf S}$ is determined by the angular-velocity projection (see Fig.~\ref{Fig_Coriolis})
\begin{equation}
\label{angular_velocity}
{{\bm\Omega}_\tau } \cdot {\bf{\bar s}} = \frac{{ - {\bf{A}} \cdot { \dfrac{{d{\bf{B}}}}{{d\tau }}}}}{{2 \left| {\bf{A}} \right|\left| {\bf{B}} \right|}} 
= - {\bf{\bar a}} \cdot \frac{{d{\bf{\bar b}}}}{{d\tau }}.
\end{equation}
Substituting Eqs.~(\ref{spin}) and (\ref{angular_velocity}) into Eq.~(\ref{Coriolis}), and considering $\tau$ as a parameter along the integration contour, $d\tau {\displaystyle\frac{d}{{d\tau }}}  = d{\bf{r}} \cdot \nabla$, we recover the geometric-phase expression (\ref{geometric_phase_3D}). \CKB{The above consideration shows that the geometric phase can be understood as a {\it dynamical} phenomenon, which is described by the corresponding coupling term in the Lagrangian or Hamiltonian of the system \cite{Kuratsuji1985,Kuratsuji1986,Mathur1991,Niu1996,Niu1999,
Bliokh2004,Bliokh2004II,Bliokh2008NP,Bliokh2009,Berard2006,Duval2006}.}

The PB geometric phase (\ref{PB_phase}) for paraxial 2D fields readily follows from the Coriolis-Doppler expression (\ref{Coriolis}). Indeed, in the paraxial regime, the spin is directed along the $z$-axis, ${\bf S} = S\, \bar{\bf z}$, and it equals the third Stokes parameter (\ref{Poincare}) \CMRD{\cite{Berry1998,Dennis2004}}: $S = {{\cal S}_3} = \cos \theta$, while the orientation of the polarization ellipse in the $(x,y)$-plane is given by half the azimuthal angle $\phi$ on the Poincar\'{e} sphere, i.e., the angular velocity of the ellipse rotation is also aligned with the $z$-axis and equals ${\bm{\Omega}_\tau } = {\displaystyle \frac{1}{2}} {\displaystyle \frac{d\phi}{d\tau}}\, \bar{\bf z}$. The product ${\bf S} \cdot {\bm{\Omega}_\tau }\, d\tau  = {\displaystyle \frac{1}{2}}\cos \theta \,d\phi$ immediately yields the Poincar\'{e}-sphere equation (\ref{PB_phase}). In terms of the polarization-ellipse vectors $\bar{\bf a}$ and ${\bar{\bf b}}$, 
\CMRD{the following relations hold:}
\begin{equation}
\label{ab}
{\bf{\bar a}} = {\bf{\bar x}}\cos \frac{\phi }{2} + {\bf{\bar y}}\sin \frac{\phi }{2}~,~~
{\bf{\bar b}} =  \sigma\!\left(- {\bf{\bar x}}\sin \frac{\phi }{2} + {\bf{\bar y}}\cos \frac{\phi }{2}\right),
\end{equation}
where $\sigma={\rm sgn}(S)={\rm sgn}(\cos\theta)$ determines the handedness of the polarization. This leads to ${\bf{\bar b}} \cdot d{\bf{\bar a}} =  - {\bf{\bar a}} \cdot d{\bf{\bar b}} = \sigma\, d\phi /2$.

\begin{figure}
\centering
\includegraphics[width=0.65\linewidth]{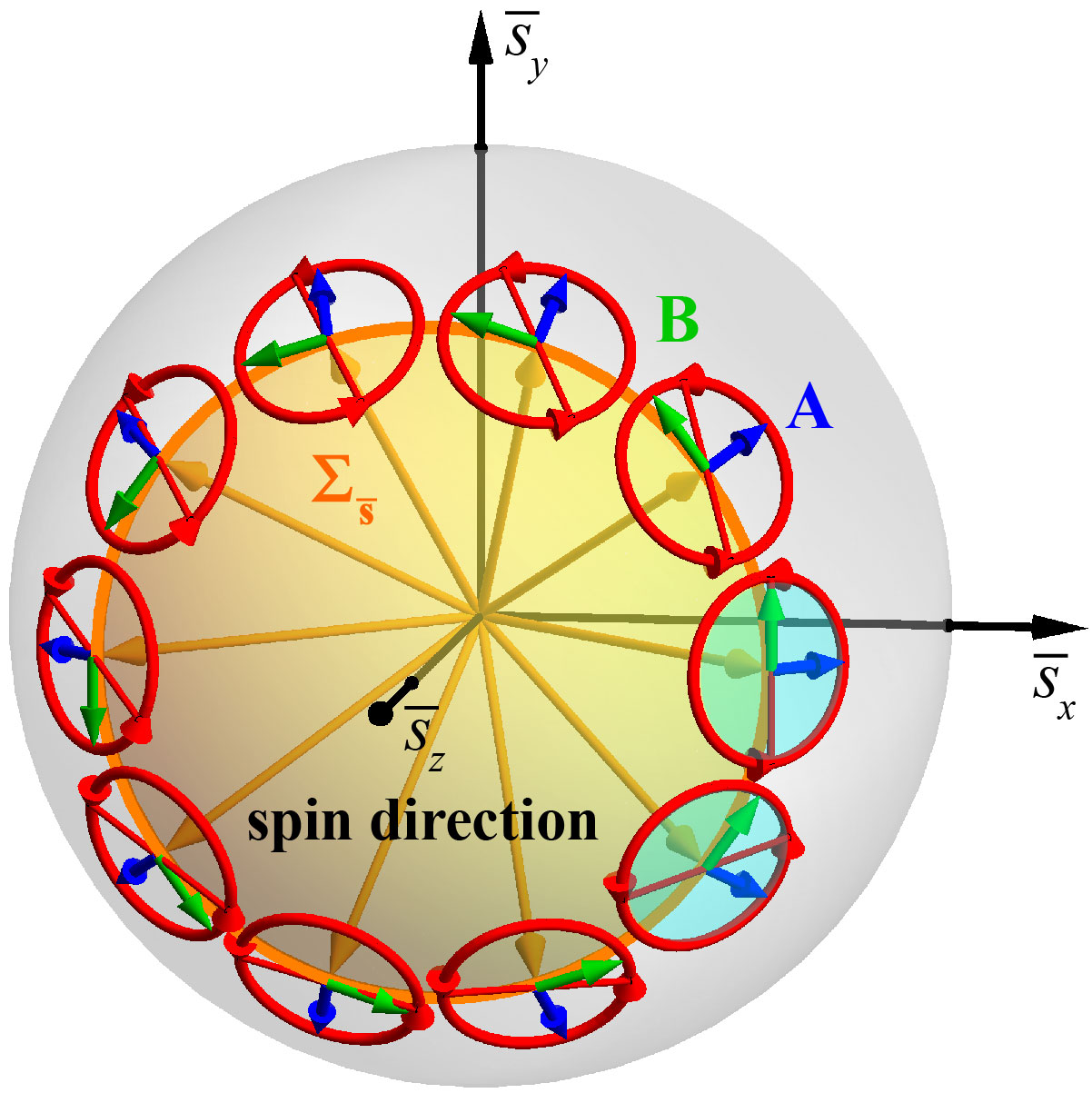}
\caption{\CKB{Unit sphere of spin directions $\bar{\bf s}$. \CMRD{The cyclically}-varying direction of circularly-polarized waves traces a closed contour on this sphere. Polarizations circles and their vectors $({\bf A},{\bf B})$ (defined up to an arbitrary rotation) are tangent to this sphere and shown here \CMRD{on} a much smaller scale. The spin-redirection geometric phase (\ref{spin_redirection}) and (\ref{spin_redirection2}) is indicated by red diameter lines inside the \CMRD{polarization} circles, which follow the parallel-transport law on the sphere. The initial and final parallel-transported polarizations are highlighted in cyan. \KB{The angle between the initial and final red diameter lines} equals the geometric phase ${\Phi _{\rm G}}\,{\rm mod}\,2\pi = \Sigma_{\bar{\bf s}}$.}}
\label{Fig_spin}
\end{figure}

Thus, when the direction of the spin ${\bf S}$ is fixed (up to the sign) whereas its {\it magnitude} $S$ can vary, the 3D geometric phase (\ref{Coriolis}) reduces to the PB phase. In \CKB{the opposite} case, when the magnitude of the spin is fixed and maximal, $|{\bf S}|=\sqrt{2}\, |{\bf A}|=\sqrt{2}\, |{\bf B}|=1$ (i.e., the field is circularly polarized), whereas its {\it direction} $\bar{\bf s}$ varies, the Coriolis equation (\ref{Coriolis}) provides the {\it spin-redirection} (or Bortolotti-Rytov-Vladimirskii-Berry) phase \cite{Bortolotti,Rytov,Vladimirskii,Ross1984,Chiao1986,Tomita1986,Berry1987_II,
Haldane1986,Segert1987,Lipson1990,BB1987,Jordan1987}.
This geometric phase can be written as
\begin{equation}
\label{spin_redirection}
{\Phi _{\rm G}} = - \int {{\bm{\Omega}}_\tau \cdot {\bf{\bar s}}} \;\, d\tau  
 = \oint {{\bf{\bar a}} \cdot d{\bf{\bar b}}}~. 
\end{equation}
Introducing the {\it sphere of spin directions} determined by the unit spin vector $\bar{\bf s}=(\bar{s}_x,\bar{s}_y,\bar{s}_z)$, Fig.~\ref{Fig_spin}, the phase (\ref{spin_redirection}) modulo $2\pi$ is numerically equal to the oriented area (solid angle) enclosed by the contour on this sphere
(see \cite{Haldane1986,Segert1987,Bliokh2009} for the derivation of this fact using a triad of mutually orthogonal vectors).  
This is similar to the PB phase on the Poincar\'e sphere, but without \CMRD{both the factor of $1/2$ and} the peculiar singular behaviour near polarization singularities. 
(Indeed, singularities of the spin direction are ``L-lines'' of linear polarization \CMRD{\cite{Dennis2009,Nye1987,BerryDennis2001}}, but we now consider purely-circularly-polarized fields.) 

\CKB{It is important to note the following peculiarities of the spin-redirection phase (\ref{spin_redirection}). First, the vectors $(\bar{\bf a},\bar{\bf b})$ are not uniquely defined for circular polarizations; these vectors can be rotated by an arbitrary angle with respect to $\bar{\bf s}$
\CMRD{that directly adds to the (arbitrary) vibration phase $\epsilon$}.
Therefore, for closed contours, where the initial and final points have the same choice of $(\bar{\bf a},\bar{\bf b})$, the spin-redirection phase (\ref{spin_redirection}) can differ by an integer times $2\pi$, depending on the choice of the continuously-varying $(\bar{\bf a},\bar{\bf b})$ vectors for all circular polarizations along the contour of evolution. For example, choosing the $(\bar{\bf a},\bar{\bf b})$ aligned with unit vectors of spherical coordinates $(\vartheta,\varphi)$ on the $\bar{\bf s}$-sphere, as shown in Fig.~\ref{Fig_spin}, the geometric phase (\ref{spin_redirection}) can be written as \cite{Bliokh2009}
\begin{equation}
\label{spin_redirection2}
{\Phi _{\rm G}} = - \oint {\cos\vartheta\, d{\varphi}}, \quad
{\Phi _{\rm G}}\,{\rm mod}\,2\pi = \Sigma_{\bar{\bf s}},
\end{equation}
where $\Sigma_{\bar{\bf s}}$ is the oriented spherical area (solid angle) enclosed by the contour. This freedom in the choice of the $(\bar{\bf a},\bar{\bf b})$ vectors  corresponds to the gauge freedom for the Berry connection on the sphere of directions \cite{Bliokh2009,Bliokh2010}, and it distinguishes the spin-redirection geometric phase from the general case, where the polarization-ellipse vectors, and hence the geometric-phase value (not modulo $2\pi$), are uniquely defined. Second, we note that the geometric phase (\ref{spin_redirection}) corresponds to the {\it parallel transport} of  circular polarizations on the sphere of directions \cite{Shapere_book,Bortolotti,Rytov,Vladimirskii,Ross1984,Chiao1986,Tomita1986,Berry1987_II,
Haldane1986,Segert1987,Vinitskii1990,Bhandari1997,Ben-Aryeh2004,Malykin2004,Bliokh2009}. Denoting the geometric phase of the circular polarizations tangent to the $\bar{\bf s}$-sphere by vectors (see red lines inside polarization circles in Fig.~\ref{Fig_spin}), one can see that these vectors follow the geometric parallel transport on the sphere. Equivalently, if we would introduce vectors $(\bar{\bf a},\bar{\bf b})$ parallel-transported on the sphere, the phase (\ref{spin_redirection}) would vanish. 
However, such vectors cannot be \CMRD{globally} defined without discontinuities along a generic closed contour.}

We finally note that the spin-redirection geometric phase represents a {\it highly-degenerate case} in our general treatment of inhomogeneous polarized fields. Indeed, it corresponds to a ``global C-point'' spread over the whole contour of evolution. 
In fact, the field $\Psi={\bf E}\cdot {\bf E}$ vanishes \CMRD{everywhere} in a purely circularly-polarized field, and one cannot unambiguously separate the dynamical and geometric phases along a closed spatial contour. Nonetheless, the spin-redirection geometric phase does play an important role and becomes generic for inhomogeneous {\it free-space optical fields}. Although such fields are generally elliptically-polarized in real space, they can be considered as Fourier superpositions of circularly-polarized plane waves with different wavevectors and helicities \CMRD{\cite{BerryDennis2001,Bliokh2010}}. Indeed, circularly-polarized plane waves are the helicity-momentum eigenstates of Maxwell equations \cite{BLP}. Due to the transversality of electromagnetic plane waves, $\nabla \cdot {\bf E} = {\bf k} \cdot {\bf E} =0$, their spin direction is locked with the direction of the wavevector, $\bar{\bf s} = \sigma \bar{\bf k}$ (where $\sigma=\pm 1$ is the helicity), and the spin-direction sphere becomes a {\it sphere in momentum ${\bf k}$-space} \cite{BB1987,Jordan1987,Skagestam1992,Bliokh2010}. Therefore, the spin-redirection construction, highly-degenerate in terms of real-space polarizations, becomes general and exact in terms of the {\it momentum representation} of free-space Maxwell fields.
This is why the Berry connection and curvature associated with the ${\bf k}$-space sphere and spin-redirection geometric phase appear in relativistic wave equations for massless spinning particles (e.g., photons) on a very fundamental level \cite{BB1987,Jordan1987,Skagestam1992} and determine position, spin and orbital angular momentum operators for such particles \cite{Bliokh2010,Berard2006}. However, this approach becomes approximate for optical fields in isotropic smoothly-inhomogeneous media \cite{Rytov,Vladimirskii,Ross1984,Chiao1986,Tomita1986,Berry1987_II,Bliokh2008NP,
Bliokh2009} and generally inapplicable in sharply-inhomogeneous or anisotropic media, where helicity is not conserved and circularly-polarized plane waves are not eigenmodes of the \KBII{problem.}

Coming back to generic 3D polarized fields and the general geometric phase (\ref{geometric_phase_3D}) and (\ref{Coriolis}), it is valid in fields with simultaneously varying spin direction and magnitude. Therefore, it combines features of the PB and spin-redirection geometric phases. To characterize the geometrical and topological properties of this general geometric phase, we need a more sophisticated spherical representation, combining the main features of the Poincar\'e and spin-direction spheres. We describe this new sphere, as well as topological properties of polarization singularities on 3D fields, in the next subsection.  

\subsection{\normalsize ``Poincarana-sphere'' representation and polarization M\"obius strips}
\label{Sec-Poincarana}

As discussed in Section~\ref{Pancharatnam}, only two parameters (e.g.\CMRD{,} the eccentricity and the orientation of the polarization ellipse) are required to characterize the polarization state of a fully polarized paraxial 2D field, and these parameters can be represented by a point over the Poincar\'e sphere, Fig.~\ref{Fig_Poincare}.
One way of understanding this sphere is by using the stereographic projection: the (possibly infinite) complex scalar \CMRD{$E^-/E^+$} is equated with a stereographic variable $\zeta = \tan(\theta/2) e^{i \phi}$, implying $\alpha^- - \alpha^+ = 2\phi$, Eqs.~(\ref{2D_field}) and (\ref{Poincare}).

\begin{figure*}[!t]
\centering
\includegraphics[width=0.85\linewidth]{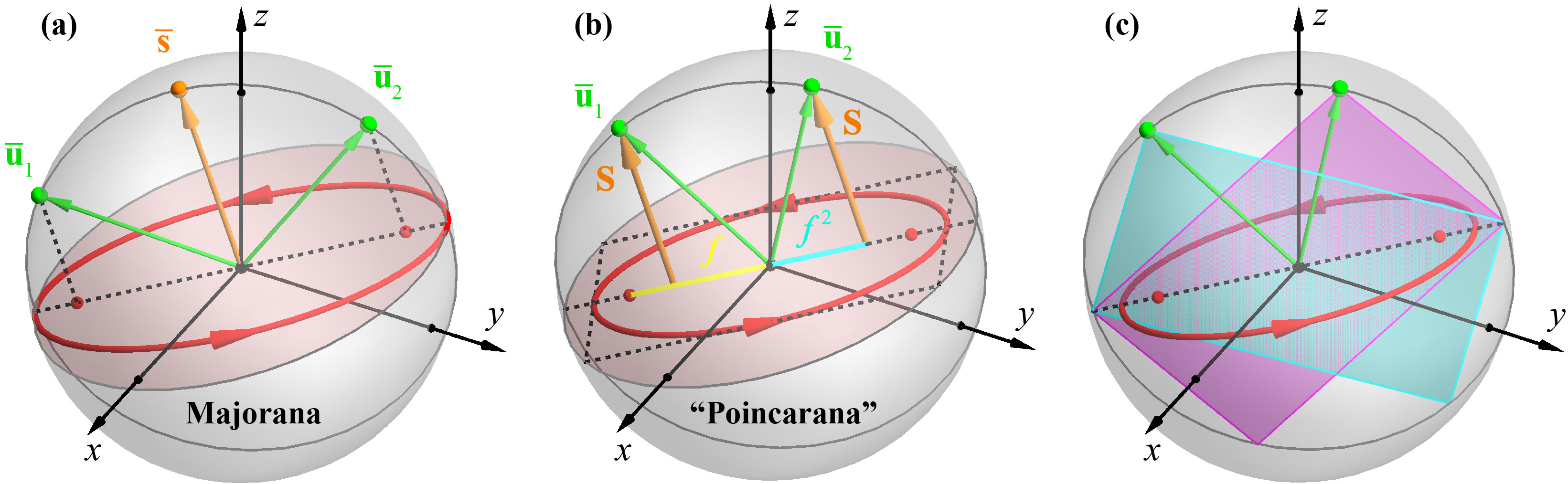}
\caption{\CKB{The Majorana (a) and ``Poincarana'' (b,c) spheres for 3D polarized fields. The 3D polarization ellipse, Fig.~\ref{Fig_Coriolis}, is normalized and placed inside the unit sphere in real space. The representation points (unit vectors) $\bu_{1,2}$ on the spheres, Eqs.~(\ref{upm}) and (\ref{upmPoincarana}), are obtained by projecting the focal (Majorana) and squared-focal (Poincarana) points of the ellipse onto the sphere in the normal direction of spin ${\bf S}$. 
The Majorana and Poincarana representations differ by the normalization of the polarization ellipse with respect to its major semiaxis (Majorana) and energy $|{\bf A}|^2+|{\bf B}|^2$ (Poincarana). 
In the latter case, the rectangles with corners at ${\bf{\bar a}},{\bf{\bar u}}_1,-{\bf{\bar a}},-{\bf{\bar u}}_2$ (cyan) and ${\bf{\bar a}},{\bf{\bar u}}_2,-{\bf{\bar a}},-{\bf{\bar u}}_1$ (magenta) shown in (c) are rotated versions of the rectangle including the polarization ellipse and shown by the dashed line in (b).}}
\label{Fig_Poincarana}
\end{figure*}

The Poincar\'e sphere, based on the two-component ``wavefunction'' (\ref{2D_field}), is analogous to the Bloch sphere for spin-1/2  particles, where a general state $a\, |\!\uparrow\,\rangle + b\, |\! \downarrow \, \rangle$ corresponds to a state in the stereographic $\zeta$ direction with $\zeta = -b/a$, or equivalently the solution of the linear polynomial $a\, \zeta + b = 0$. \CKB{(It should be emphasized, however, that the Bloch sphere represents the actual {\it spin angular momentum in real ${\bf r}$-space}, while the Poincar\'e sphere is the representation of 
\CMRD{the} Stokes vector, 
sometimes called ``pseudospin'', in the abstract $\vec{\mathcal{S}}$-space of the Stokes parameters.)}  
In 1932, Majorana \cite{Majorana1932} generalized this spherical representation to an arbitrary quantum spin $n=1/2,1,3/2,...$, i.e., for \CMRD{the} $(2n+1)$-dimensional \CMRD{irreducible} representations of \CMRD{the group SU(2)}. According to his approach, {\it a general spin-$n$ state $\sum_{i = -n}^n a_j |j\rangle$ is represented by $2n$ points on the unit ${\rm S}^2$ sphere}; these points can be found as the stereographic coordinates of the roots of the polynomial of order $2n$: $\sum_{j = -n}^n (-1)^j \left( \begin{smallmatrix}2j \\ j+n\end{smallmatrix}\right)^{1/2} a_j\, \zeta^{j+n}$\CKB{, where $\left( \begin{smallmatrix}2j \\ j+n\end{smallmatrix}\right)$ is the binomial coefficient}.
The points on the {\it Majorana sphere} cannot be \CMRD{labelled in a global, unambiguous way}: around a closed circuit, they may undergo a permutation.
{The Majorana sphere has been used to describe quantum aspects of light's polarization \cite{Bjork2015}, 
\CMRD{geometric properties of multipoles \cite{mrd2004},} 
and even to characterize a class of optical beams \cite{GutierrezAlonso,GutierrezDennisAlonso}.
\CMRD{In the spinorial approach to relativistic tensors, the Majorana spin directions are called ``principal null directions'' \cite{PenroseRindler1,PenroseRindler2}.}

\CMRD{Using the Majorana representation, Hannay found the general formula for the geometric phase on a closed circuit accumulated by such a general spin-$n$ state \cite{Hannay1998_II}}. 
\CKB{Since electromagnetic waves correspond to spin-1 particles (photons) with the three-component ``wavefunction'' ${\bf E}$, the Majorana representation for optical fields should involve {\it two points} on the unit sphere.} Hannay considered this spin-1 case \cite{Hannay1998} and found a simple geometric interpretation of the two points (unit vectors) $\bu_{1,2}$ on the Majorana sphere in terms of the general 3D polarization ellipse (\ref{AB}), Fig.~\ref{Fig_Poincarana}(a)\CMRD{: } 
\CKB{the unit vector directed along the bisector} of the two vectors $\bu_{1,2}$ is the direction normal to the polarization ellipse, i.e., the unit spin vector $\bar{\bf s}$. At the same time, the projections of the two points $\bu_{1,2}$ in the direction of $\bar{\bf s}$ onto the plane of the polarization ellipse give the locations of the two foci of the polarization ellipse, whose major semiaxis is normalized to unity, see Fig.~\ref{Fig_Poincarana}(a). 
 
\CMRD{A geometric interpretation of the Majorana representation for light} is that the projection of the polarization ellipse onto either of the directions $\bar{\bf u}_{1,2}$ gives a circle whose rotation follows the right-hand rule \CMRD{\cite{penrosemind}}.
This property connects it directly with measurement techniques proposed recently \cite{Nechayev2018} in which the local 3D polarization is probed by placing a nanoparticle at the point in question and the scattered field is collected by a microscope objective, so that the 3D polarization is retrieved by the position and handedness of two far-field C-points within a hemisphere, which essentially correspond to $\pm\bar{\bf u}_1$ and $\pm\bar{\bf u}_2$, where the signs depend on the C-points' helicity.
\CMRD{This property generalizes to all other spins and can be considered the geometric definition of the Majorana sphere  \cite{mrd2004,PenroseRindler1,PenroseRindler2}.}
 
Thus, the Majorana sphere for spin 1 describes the polarization ellipse with \emph{any} orientation \CMRD{using} four parameters. 
\CKB{This sphere can be considered as a unification of the Poincar\'e and spin-direction spheres, Figs.~\ref{Fig_Poincare} and \ref{Fig_spin}, because it simultaneously represents the direction of the spin (two parameters),
\CMRD{represented by $\bar{\bf s}$, parallel to $\bu_1 + \bu_{2}$,}
 and the properties of the polarization ellipse in the polarization plane (the other two parameters),}
\CMRD{represented by $\bu_1 - \bu_{2}$.} 
When the ellipse is known to be confined to the $(x,y)$ plane, \CKB{the spin-direction} parameters become redundant, and the Majorana sphere can be mapped onto the Poincar\'e sphere \CMRD{\cite{Hannay1998}}.
\CKB{\CMRD{It is important to note} the following fundamental difference between the Poincar\'e and Majorana representations. The Poincar\'e sphere, Fig.~\ref{Fig_Poincare}, is a unit sphere {\it in the abstract \CMRD{Stokes-parameter} space}, where the vector $\vec{\mathcal{S}}$ does not represent the actual spin of the electromagnetic field. 
\CMRD{As stated previously,} only the third component of this vector is related to the $z$-directed spin angular momentum: ${\bf S}={\mathcal{S}}_3 \bar{\bf z}$ \CMRD{\cite{Berry1998,Dennis2004}}. In contrast to this, the Majorana sphere is a unit sphere {\it in real ${\bf r}$-space}, and the vectors $\bu_{1,2}$ indicate the direction of the actual spin angular momentum of the field, $\bar{\bf s}$, as shown in Fig.~\ref{Fig_Poincarana}(a). Since the geometric phase of a generic 3D field is closely related to this spin via the Coriolis representation (\ref{Coriolis}), one could expect that the Majorana sphere would serve as an effective geometric tool for the geometric-phase calculations.}

However, a drawback to the Majorana sphere is that, unlike the Poincar\'e sphere, it does not geometrically incorporate the {\it normalization} of the polarization ellipse: the polarization ellipse given by the construction above is normalized with respect to its semimajor axis $|{\bf E}||{\bf A}|$ rather than its intensity $|{\bf{E}}|^2$. 
\CMRD{Any convenient representation of the geometric phase must incorporate this intensity normalization. In the following, we propose an alternative Majorana-like representation \CMRD{that naturally incorporates the correct normalization}, and hence the geometric phase becomes easier to interpret.} 

Consider a generalization of the Majorana construction described above, in which the unit vectors $\bu_{1,2}$ are bisected by $\bar{\bf s}$ and whose separation $\bu_1-\bu_2$ is parallel to the major polarization-ellipse axis $\bar{\bf a}$, following the expression
\begin{align}
\bu_{1,2}=\pm\sqrt{1-\beta^2}\,\,\bar{\bf a}+\beta\,\bar{\bf s},
\label{upm}
\end{align}
where $\beta$ parametrizes the eccentricity of the polarization ellipse. The standard Majorana-sphere representation proposed by Hannay uses $\beta=|{\bf B}|/|{\bf A}|$. However, as shown in Appendix~\ref{apppoincarana}, choosing instead $\beta=|{\bf S}|=2|{\bf A}||{\bf B}|$ 
leads to a direct connection with the geometric phase in terms of the enclosed solid angle on the sphere (similar to the PB and spin-redirection geometric phases). With this choice, Eq.~(\ref{upm}) reduces to
\begin{align}
\bu_{1,2} =\pm\sqrt{1-|{\bf S}|^2}\,\,\bar{\bf a}+{\bf S}
=\pm f^2\,\bar{\bf a}+{\bf S},
\label{upmPoincarana}
\end{align}
where $f=\sqrt{|{\bf A}|^2-|{\bf B}|^2}$ is the focal distance of the polarization ellipse with normalized intensity, i.e., with $|{\bf A}|^2+|{\bf B}|^2=1$, as was used in Section~\ref{Sec-Coriolis} and shown in Fig.~\ref{Fig_Poincarana}(b). 
Remarkably, for paraxial light with the polarization ellipse lying in the $(x,y)$ plane, the $z$-components of the two representation points coincide with the height of the point on the Poincar\'e sphere, and hence with the spin magnitude: 
$\bar{u}_{1,2z} = {\cal S}_3 = S_z$. Thus, this representation mixes features from the Poincar\'e and the Majorana spheres, and we call it the {\it ``Poincarana'' sphere} representation, Fig.~\ref{Fig_Poincarana}(b,c). \CKB{Note that for purely circularly-polarized fields, the two points on the Poincarana and Majorana spheres merge, $\bu_1=\bu_2=\bar{\bf s}$, and these spheres reduce to the spin-direction sphere, Fig.~\ref{Fig_spin} \CMRD{\cite{mrdthesis}}.}

\CKB{Appendix~\ref{apppoincarana} gives a general proof that the Poincarana-sphere choice $\beta=|{\bf S}|$ provides a direct connection between enclosed solid angles and geometric phase. In the most general case, {\it the geometric phase (\ref{geometric_phase_3D}) and (\ref{Coriolis}) is equal to half the oriented area (solid angle) swept on the Poincarana sphere by the shortest geodesic line connecting \KB{the points $\bu_1$ and $-\bu_2$ (or equivalently $-\bu_1$ and $\bu_2$)}}, Fig.~\ref{Lambert}(a):
\begin{equation}
\Phi_{\rm G} = \frac{1}{2}\,\Sigma_{\rm geod}.
\label{Phase_Poincarana}
\end{equation}
It is easy to trace the transition of this equation to the PB geometric phase in the paraxial 2D case. Since the third (vertical) coordinates of the representation points on the Poincar\'e and Poincarana spheres are the same in this case, the geodesic line connecting $\bu_1$ and $-\bu_2$ on the Poincarana sphere exactly doubles the shortest geodesic line from $\vec{\cal S}$ to the equator of the Poincar\'e sphere, Eq.~(\ref{PB_phase}) and Figs.~\ref{Fig_Poincare},\ref{phasejump}. Taking into \CMRD{account} that the rate of change of the azimuthal coordinate on the Poincar\'e sphere is exactly twice  that for the Poincarana sphere (representing the orientation angle $\phi/2$ of the polarization ellipse), we find that the geodesic area $\Sigma_{\rm geod}$ on the Poincarana sphere is exactly equal to the area swept by geodesic to the equator $\Sigma_{\rm equat}$ on the Poincar\'e sphere. This precisely agrees with the definition (\ref{PB_phase}) of the PB phase.}

\begin{figure}[!t]
\centering
\includegraphics[width=1\linewidth]{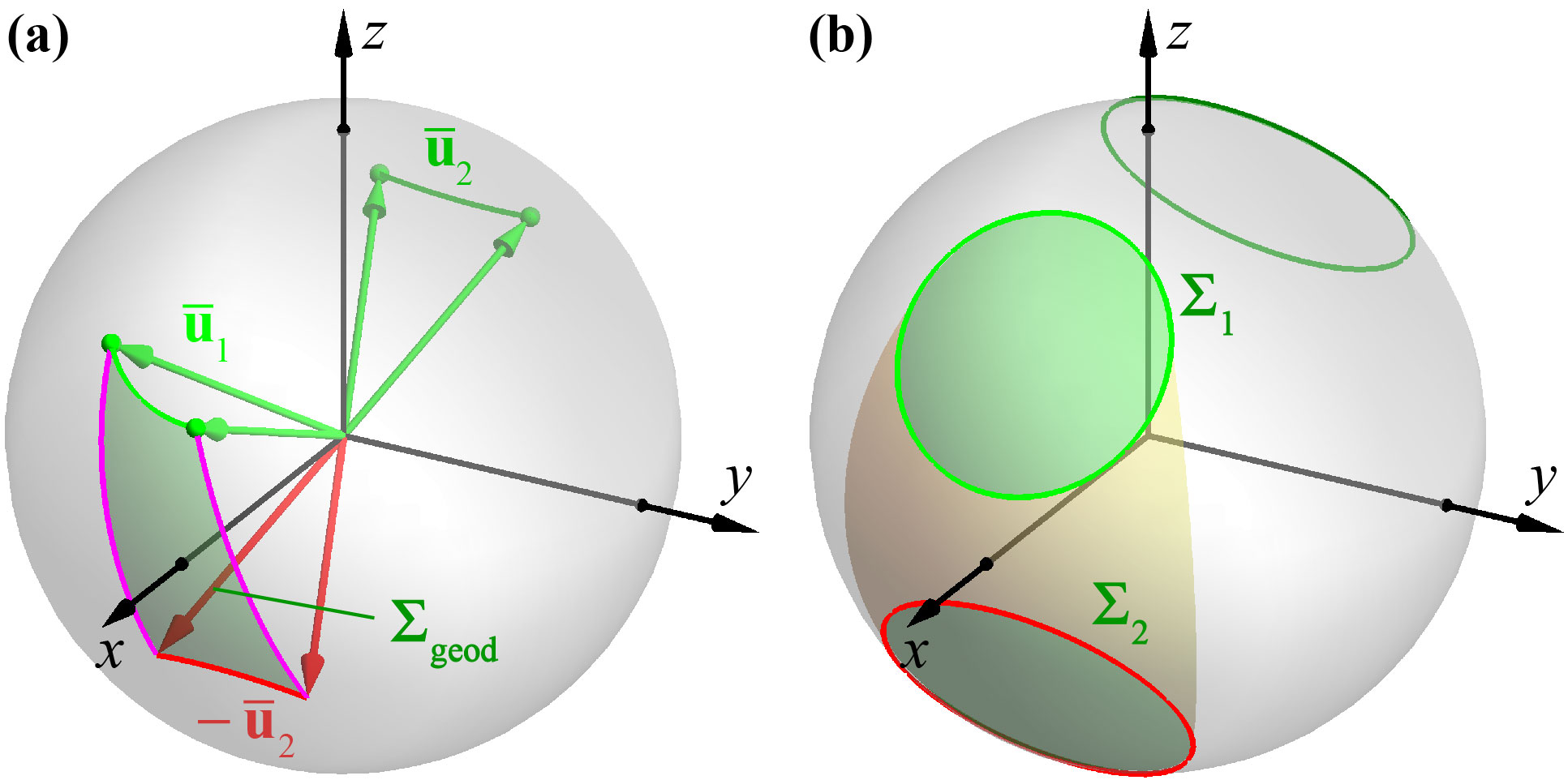}
\caption{\CKB{The oriented areas on the Poincarana sphere determining the 3D geometric phase (\ref{Phase_Poincarana}) and (\ref{PhG1}), cf. 2D case in Figs.~\ref{Fig_Poincare} and \ref{phasejump}. (a) The area $\Sigma_{\rm geod}$ is swept by the shortest geodesic line (shown in magenta) connecting points $\bu_1$ and $-\bu_2$. (b) When the points $\bu_{1,2}$ trace two closed contours on the Poincarana sphere, the area $\Sigma_{\rm geod}$ modulo $4\pi$ is equal to the sum of the oriented areas $\Sigma_{1,2}$ enclosed by the two loops. The area between the contours, highlighted by yellow, is swept twice in opposite directions and hence does not contribute. }}
\label{Lambert}
\end{figure}

\CKB{In the simplest case in which the area inside the spatial contour of integration does not include polarization singularities (considered below) and the contours traced by the $\bu_1$ and $-\bu_2$ points on the Poincarana sphere are two separated loops, Fig.~\ref{Lambert}(b), the oriented area $\Sigma_{\rm geod}$ modulo $4\pi$ becomes equivalent to the sum of the oriented areas \CMRD{enclosed} by the two loops: $\Sigma_{\rm geod} = \Sigma_1 + \Sigma_2$. This is similar to the paraxial PB case shown in Fig.~\ref{phasejump}(a). }

\CKB{We are now in a position to consider the role of {\it polarization singularities} of the 3D fields on the geometric phase. Similarly to the 2D case, the phase singularities of the field $\Psi = {\bf E}\cdot {\bf E}$ are {\it C-lines} of purely circular polarization in the $(x,y,z)$-space or C-points in the $(x,y)$-plane \cite{Nye1987,BerryDennis2001,Dennis2009}. Note, however, that the orientation of the normal to these circular polarizations can be arbitrary, and singularities of the full 3D field $(E_x,E_y,E_z)$ do not coincide with singularities of its $(E_x,E_y)$ transversal components (unless $E_z=0$) \cite{Nye1987,BerryDennis2001,Dennis2009}. 
Most importantly, in 2D fields considered in Section~\ref{Pancharatnam}, we characterized polarization singularities 
by the integer topological index $N_{\rm C}$, Eqs.~(\ref{C_phase_2D}) and (\ref{C_charge_2D}), counting the number of half-turns of the 2D polarization ellipse around the C-point. However, {\it this topological number cannot be generalized to the 3D case}. Indeed, the number of turns makes sense for a planar object undergoing Abelian SO(2) rotational evolution, but this does not make sense for 3D objects under non-Abelian SO(3) evolution. In terms of spherical representations, one can count the winding number of the contour around the ${\mathcal S}_3$ axis of the Poincar\'e sphere (responsible for the spin and polarization singularities), but one cannot introduce the winding number for the contours on the Poincarana/Majorana spheres because the relevant rotation of the polarization ellipse occurs with respect to the instantaneous $\bar{\bf s}$-direction, which itself evolves along the contour.  
Therefore, only the quantized dynamical phase $\Phi_{\rm D}$ (\ref{phase_3DD}) and the corresponding half-integer topological number $N_{\rm D}/2$ survive for 3D fields. 
Nonetheless, 3D polarization singularities do play an important role in the {\it topological} properties of geometric phases and their representations on the \CMRD{Majorana/}Poincarana sphere.}

To show this, note that the directions of the principal axes of the polarization ellipse are defined up to a common sign flip: $({\bf A},{\bf B}) \to (-{\bf A},-{\bf B})$. This ambiguity corresponds to the ambiguity of the \CMRD{labelling} of the two points on the Poincarana sphere: $({\bu_1},{\bu_2}) \to (\bu_2,\bu_1)$. Therefore, upon a cyclic evolution of the 3D polarization ellipse, one can distinguish two topologically different cases, shown in Fig.~\ref{areas} \cite{mrdthesis}.

{\bf Case~1:} Upon continuous cyclic evolution, the principal-axis vectors return to their initial values: $({\bf A},{\bf B})_{\rm in} = ({\bf A},{\bf B})_{\rm fin}$. This means that the vectors $\bu_1$ and $\bu_2$ trace {\it two closed loops over the Poincarana sphere}, as shown in Figs.~\ref{areas}(a,b). These paths generally have different shapes and could have different handedness, i.e., different signs of the enclosed solid angles. In this case, the geometric phase (\ref{Phase_Poincarana}) modulo $2\pi$ equals the half sum of the oriented areas of the two loops on the Poincarana sphere:
\begin{subequations}
\begin{equation}
\Phi_{\rm G}\, {\rm mod}\, 2\pi=\frac{1}{2}\left(\Sigma_1 + \Sigma_2\right) {\rm mod}\, 2\pi.
\label{PhG1}
\end{equation}

{\bf Case~2:} 
Upon continuous cyclic evolution, the polarization-ellipse vectors return to the values opposite to the initial ones: $({\bf A},{\bf B})_{\rm in} = (-{\bf A},-{\bf B})_{\rm fin}$. Each of these vectors, traced along the spatial contour of evolution, forms a 3D structure similar to a {\it M\"obius strip} with a half-integer number of turns around the contour \cite{Freund2010,Freund2010II,Dennis2011,Bauer2015,Galvez2017,Garcia2017,
Bauer2019}. In this case, the vectors $\bu_1$ and $\bu_2$ swap after the cyclic evolution, $({\bu_1},{\bu_2})_{\rm in} = ({\bu_2},{\bu_1})_{\rm fin}$, and form {\it a single closed loop over the Poincarana sphere}, Fig.~\ref{areas}(c). Remarkably, one can show that the geometric phase (\ref{Phase_Poincarana}) modulo $2\pi$ becomes equal to half of the solid angle enclosed by this loop, but now with {\it an additional $\pi$ contribution}:
\begin{equation}
\Phi_{\rm G}\, {\rm mod}\, 2\pi=\left(\frac{1}{2}\,\Sigma + \pi \right) {\rm mod}\, 2\pi.
\label{PhG2}
\end{equation}
\end{subequations}
%

\begin{figure}
\centering
\includegraphics[width=1\linewidth]{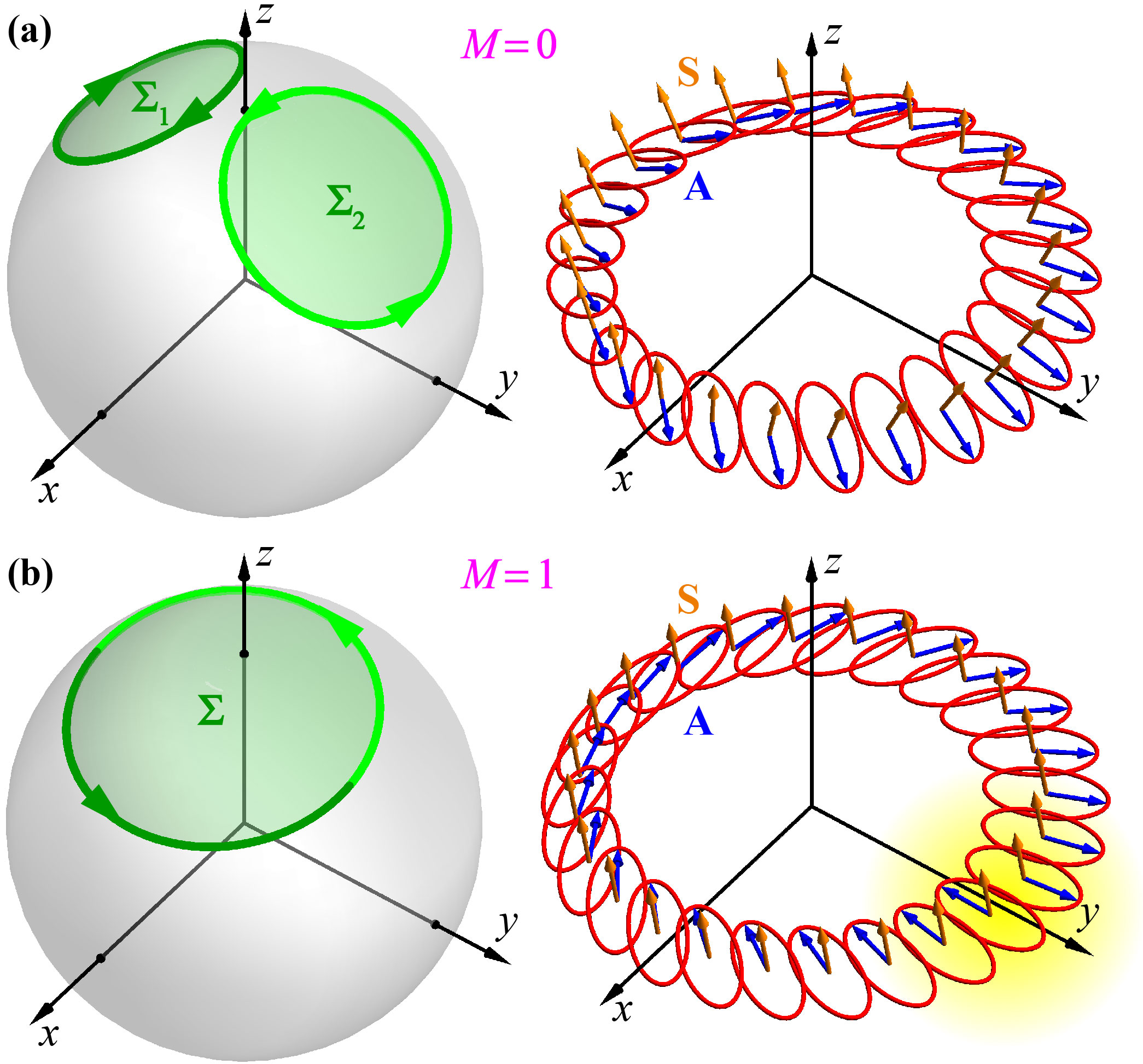}
\caption{\CKB{Two topologically-different cases of closed contours on the Poincarana sphere and the corresponding spatial 3D polarization distributions. (a) {\bf Case~1 ``non-M\"obius''}. The vectors $\bu_{1,2}$ trace two closed loops on the Poincarana sphere. The major semiaxis ${\bf A}$ of the polarization ellipse experiences a continuous 3D evolution along the spatial contour. (b) {\bf Case~2 ``M\"obius''}.
The vectors $\bu_{1,2}$ trace a single closed loop exchanging their indices $1,2$ at some point (which can be chosen arbitrarily on the contour). The corresponding spatial distribution of  the major semiaxis of the polarization ellipse has a discontinuity ${\bf A} \to -{\bf A}$ highlighted in yellow. This forms the polarization  M\"obius strip \cite{Freund2010,Freund2010II,Dennis2011,Bauer2015,Galvez2017,Garcia2017,
Bauer2019}. The transition between the two cases occurs each time that the spatial contour crosses a non-degenerate C-line. The two cases are marked by the $\mathbb{Z}_2$ topological number $M$, Eq.~(\ref{Z2}).}}
\label{areas}
\end{figure}

\begin{figure*}
\centering
\includegraphics[width=0.9\linewidth]{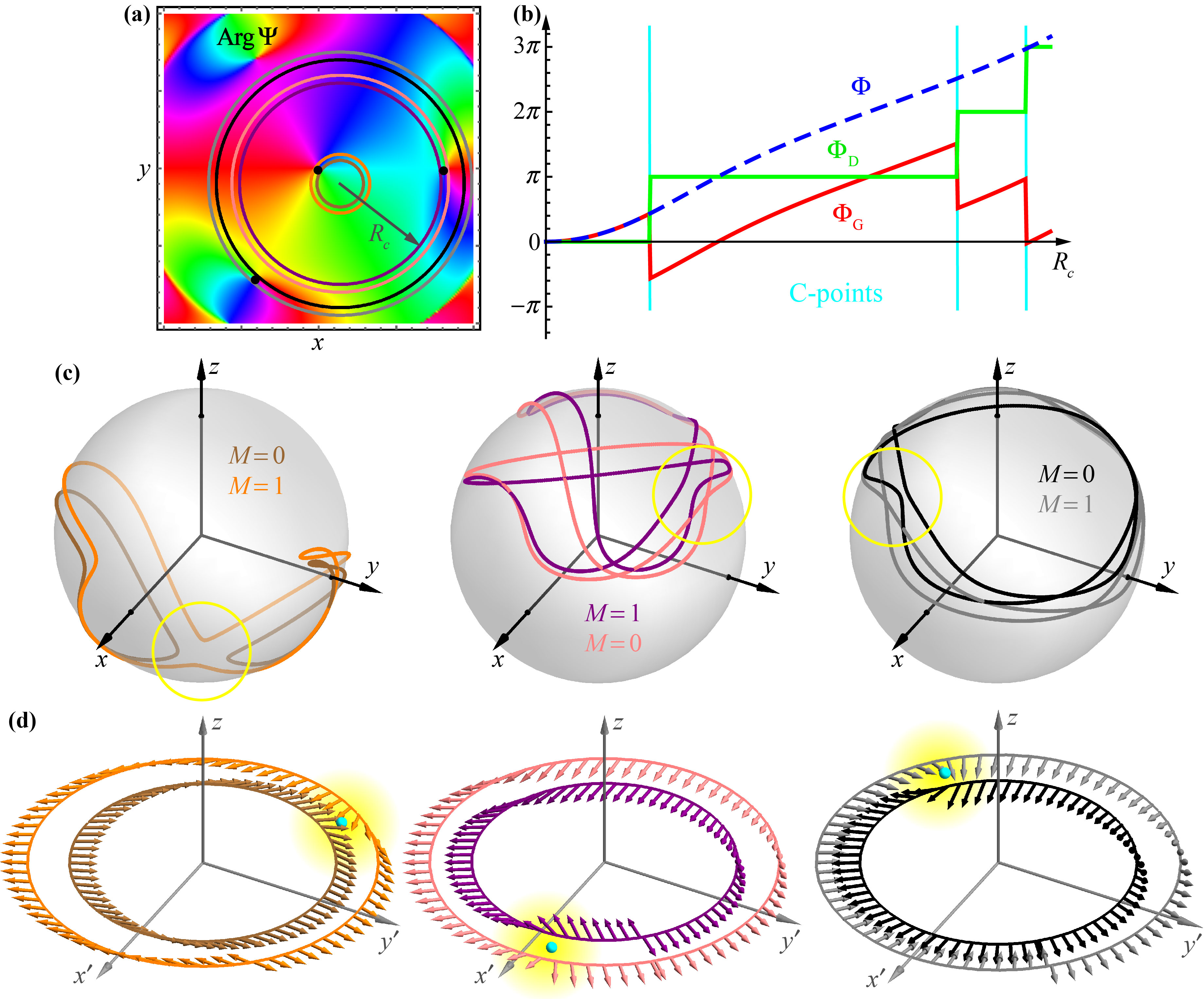}
\caption{An example of polarization singularities, M\"obius strips, and various phases in a 3D polarized field, corresponding to a focused 2D polarization vortex similar to Fig.~\ref{FPB}(a) and described by Eqs.~(\ref{focusedFP})--(\ref{Gaussian}). (a) The phase distribution of the quadratic field $\Psi={\bf E}\cdot{\bf E}$ in the waist $(x,y)$ plane indicates the C-points of the 3D field ${\bf E}$ (black dots). Three pairs of concentric circular contours correspond to crossings of three C-points. (b) The total ($\Phi$), dynamical ($\Phi_{\rm D}$), and geometric ($\Phi_{\rm G}$) phases on the radius of the contour $R_c$, cf. Figs.~\ref{FPB}(a,c). (c) Contours on the Poincarana sphere corresponding to the three pairs of spatial contours in (a). The topological $\mathbb{Z}_2$ number $M$ indicates the non-M\"obius ($M=0$) and M\"obius  ($M=1$) cases (see Fig.~\ref{areas}), whereas the yellow circles highlight the regions of the reconnection of the contours. (d) The 3D distributions of the directions of the major semi-axes of the polarization ellipses, $\bar{\bf a}$, along the three pairs of spatial contours corresponding to the panels (a) and (c). Transitions between the non-M\"obius and M\"obius cases are realized by fast $\bar{\bf a} \to -\bar{\bf a}$ rotations in one of the contours in each pair in the vicinity of the C-points (shown in cyan). These fast rotations produce opposite  $-\bar{\bf a} \to \bar{\bf a}$ discontinuities in another point (arbitrarily chosen) of the contour, cf. Fig.~\ref{areas}(b).}
\label{FPB}
\end{figure*}

\CKB{Equations (\ref{PhG1}) and (\ref{PhG2}) provide the 3D generalization of the 2D equation (\ref{area}) for the PB geometric phase. One can show that spatial contours not enclosing C-lines always correspond to Case 1, while crossing a non-degenerate C-line always produces a transition between Cases 1 and 2. Again, the principal difference between the 2D and 3D situations is that we cannot count the number of polarization-ellipse turns in the 3D case. Even though one can count the number of twists of the vectors $({\bf A},{\bf B})$ around the spatial contour, {\it this number is not topological}, i.e., can vary by an integer upon small deformations of the contour without crossing the C-lines. For example, the polarization M\"obius strips generically have orders $\pm 1/2$ or $\pm 3/2$ around a single non-degenerate C-line \cite{Freund2010,Freund2010II,Dennis2011}, and this number depends on the shape of the contour or fine non-topological properties of the field in the vicinity of the C-line. What does not depend on the contour shape, and hence represents a {\it topological} property, is the {\it parity} of the number of such twists \cite{mrdthesis}. The even and odd number of twists represent the non-M\"obius and M\"obius cases 1 and 2, respectively. Thus, we can introduce a $\mathbb{Z}_2$ topological number distinguishing these cases:}
\begin{equation}
M = N_{\rm D}\, {\rm mod}\, 2.
\label{Z2}
\end{equation}
\CKB{Indeed, the topological number corresponding to the dynamical phase (\ref{phase_3DD}), $N_{\rm D}$, equals 0 for contours not including C-lines and changes by $\pm1$ each time when the contour crosses a non-degenerate C-line. Hence, $M=0$ and $M=1$ indicate the non-M\"obius and M\"obius cases 1 and 2, respectively.}

\CKB{Note that Eqs.~(\ref{PhG1})--(\ref{Z2}) are perfectly consistent with the 2D equation (\ref{area}) for the PB geometric phase. One can unify all these equations in the following form:
\begin{equation}
\nonumber
\boxed{~~\Phi_{\rm G}\, {\rm mod}\, 2\pi = \left( \frac{1}{2}\,\Sigma + \pi M \right) {\rm mod}\, 2\pi~,~}
\label{GP_singularity_general}
\end{equation}
where $\Sigma$ denotes the total oriented area from all closed loops on either the Poincar\'e or Poincarana spheres, and we used Eq.~(\ref{parity}).}

To illustrate these ideas, we consider an explicit closed-form solution of the free-space monochromatic Maxwell equations, which is given by a
combination of nonparaxial vortex focused fields, defined as 
\begin{align}
{\bf E}_{\ell^+,\ell^-}({\bf r})=\hat{\bf C}_+\hat{L}_{{\rm sgn}(\ell^+)}^{|\ell^+|}U({\bf r})+\hat{\bf C}_-\hat{L}_{{\rm sgn}(\ell^-)}^{|\ell^-|}U({\bf r}),
\label{focusedFP}
\end{align}
where the operators $\hat{L}_{\pm}$ and $\hat{\bf C}_{\pm}$ convey vorticity and helicity to a scalar field, respectively, and are given by \cite{Alonso2011}
\begin{align}
\hat{L}_{\pm}&=\pm{\bf e}^\pm\cdot{\bf r}\times\nabla,\\
\hat{\bf C}_\pm&=\frac12\left({\bf e}^\pm+\frac{{\bf e}^\pm\cdot\nabla}{k^2}\nabla\mp\frac{{\bf e}^\pm\times\nabla}k\right),
\end{align}
and the scalar field $U$ is a nonparaxial generalization of a Gaussian beam with Rayleigh range $q$, given by \cite{Berry1994,Sheppard1998}
\begin{align}
U({\bf r})=2kqe^{-kq}\,\,{\rm sinc}\!\left[k\sqrt{x^2+y^2+(z-iq)^2}\right],
\label{Gaussian}
\end{align}
where $k=\omega/c$ is the wavenumber. (The paraxial limit results from considering $kq \gg 1$.) 
When $\ell^+$ and $\ell^-$ have different parity, the field in Eq.~(\ref{focusedFP}) is qualitatively similar to the focused beams used for the experimental observations of M\"obius polarization strips \cite{Bauer2015}. 

\CKB{Figure~\ref{FPB}(a) shows the phase of the quadratic field $\Psi={\bf E}\cdot{\bf E}$, over the transverse $(x,y)$ plane, corresponding to the waist ($z=0$) of the focused field in Eq.~(\ref{focusedFP}) for $\ell^+=1,\ell^-=0$, and $kq =5$. This is a nonparaxial 3D analogue of the polarization vortex shown in Fig.~\ref{vortex}(a). As can be appreciated from Fig.~\ref{FPB}(a), this field contains several C-lines, corresponding to phase singularities of $\Psi$. This figure also shows three pairs of (concentric) circular contours within the waist plane; each pair corresponds to crossing one of the C-points. 
The evolutions of the total, dynamical, and geometric phases with the radius of the contour $R_c$ are shown in Fig.~\ref{FPB}(b), cf. Figs.~\ref{vortex}(a,c). One can see that the total phase evolves continuously, as it should be in a generic case, while the dynamical and geometric phases experience opposite $\pi$ jumps when crossing the C-points. Figures~\ref{FPB}(c) and (d) show the contours on the Poincarana sphere and the corresponding real-space distributions of the directions of the polarization-ellipse semiaxes ${\bf A}$ for the three pairs of contours from the panel (a). One can trace transitions between the non-M\"obius and M\"obius cases indicated by the topological number $M=0,1$. Yellow circles in (c) highlight the areas of the reconnections of the Poincarana-sphere contours.}

\section{\normalsize Other properties}
\label{Sec-Other}

\subsection{\normalsize Relation to the angular momenta  of \\ cylindrical \KB{and q-plate} fields}
\label{Sec-AM}

Remarkably, the geometric, dynamical, and total phases, Eqs.~ (\ref{total_phase})--(\ref{geometric_phase}) and (\ref{phase_3DT})--(\ref{Coriolis}), calculated along circular contours in {\it cylindrically-symmetric} 3D optical fields are closely related to the {\it spin and orbital \CMRD{angular} momenta} of such fields \cite{Allen_book,Andrews_book,Molina2007,Franke2008,Bliokh2015PR,Bliokh2015NP}.

We first consider only the electric field ${\bf E}$ characterized by the energy density $w=\omega |{\bf E}|^2$, canonical momentum density ${\bf p} = {\rm Im} \left[{\bf E}^*\! \cdot (\nabla) {\bf E} \right]$ and spin AM density ${\bf s} = {\rm Im} \left( {\bf E}^*\! \times {\bf E} \right)$ \cite{Berry1998,Berry2009,Bliokh2013,Bliokh2015PR}, where $\omega$ is the frequency, and we omit inessential common prefactors. The orbital AM density is given by ${\bf l} = {\bf r} \times {\bf p}$, and the canonical total AM density is ${\bf j} = {\bf l} + {\bf s}$. Note now that the orbital AM density represents the local expectation value of the differential orbital AM operator with $z$-component $\hat{l}_z = (-i {\bf r} \times \nabla)_z = -i \partial/\partial\varphi$ and the ``wavefunction'' ${\bf E}$ \cite{Allen_book,Andrews_book,Bliokh2015PR}. In turn, considering the total-phase increment (\ref{total_phase}) along the circular contour $\left\{ {r = {\mathop{\rm const}\nolimits} ,\,\varphi  \in \left( {0,2\pi } \right)} \right\}$, we can write ${\rm Im}[\nabla \cdot d{\bf r}] = {\rm Re}[( -i\partial/\partial\varphi) d\varphi]$. One can see from here that the total-phase increment along the circular contour provides the $\varphi$-averaged normalized orbital AM density of the field:
\begin{equation}
\label{OAM}
\overline{\frac{\omega l_z}{w}} = \frac{\Phi}{2\pi},
\end{equation}
where the overbar stands for a $\varphi$ average.

Cylindrically-symmetric fields (such as eigenmodes of cylindrical waveguides \cite{Snyder,Marcuse}) can be conveniently presented in the circular-polarization basis as \cite{Picardi2018}
\begin{equation}
\label{cylidrical_fields}
E^\pm = F^\pm (r,z) e^{i(m \mp 1)\varphi}, \quad 
E_z = F_z (r,z) e^{im\varphi},
\end{equation}
where $m=0,\pm 1, \pm 2, ...$ is an integer total-AM quantum number. Obviously, the energy and the $z$-components of the AM densities are $\varphi$-independent for such fields, and the overbar in Eq.~(\ref{OAM}) can be omitted. Furthermore, the squared field for Eq.~(\ref{cylidrical_fields}) equals $\Psi= {\bf E}\cdot {\bf E} = 2 E^+ E^- + E_z^2 \propto e^{2im\varphi}$, and the dynamical-phase increment (\ref{dynamical_phase}) along the circular contour immediately yields $\Phi_{\rm D} = 2\pi m$. However, simple calculations for fields (\ref{cylidrical_fields}) show that the normalized $z$ component of the total-AM density is quantized and also equals $m$ \cite{Picardi2018,Partanen2018}, so we arrive at
\begin{equation}
\label{TAM}
{\frac{\omega j_z}{w}} = \frac{\Phi_{\rm D}}{2\pi} = m.
\end{equation}
Finally, it follows from Eqs.~(\ref{geometric_phase}), (\ref{OAM}), and (\ref{TAM}) that the spin AM density of the \CMRD{cylindrical} fields (\ref{cylidrical_fields}) is associated with the geometric-phase increment along the circular contour \cite{Picardi2018}:
\begin{equation}
\label{SAM}
{\frac{\omega s_z}{w}} = - \frac{\Phi_{\rm G}}{2\pi}.
\end{equation}

Thus, the quantization of the dynamical phase (\ref{dynamical_phase}) and the corresponding topological charge are closely related to the quantization of the total AM (\ref{TAM}) of the \CMRD{cylindrically}-symmetric field. 
Note that while the dynamical phase is quantized as a {\it half-integer} $N_{\rm D}/2$ times $2\pi$, the total AM (\ref{TAM}) is \CMRD{an} {\it integer}. This means that the topological $\mathbb{Z}_2$ number $M=0$, and there are no polarization M\"obius strips in cylindrically-symmetric fields. 
This is because cylindrically-symmetric fields are not generic and only {\it second-order} C-points can obey this symmetry.

Furthermore, the close \CMRD{relationship} between the spin AM and geometric phase, Eq.~(\ref{SAM}), can be understood using the Coriolis-Doppler interpretation (\ref{Coriolis}). In the cylindrically-symmetric case, the uniform $2\pi$ rotation between the natural polar and Cartesian coordinate frames (about the $z$-axis) takes place along the circular contour, so that $\int {\bm \Omega}_\tau\, d\tau = 2\pi \bar{\bf z}$, while the normalized spin (\ref{spin}) is ${\bf S} = \omega\, {\bf s}/w$. From here, Eq.~(\ref{Coriolis}) immediately leads to Eq.~(\ref{SAM}). 

\KB{The} AM-phase relations (\ref{OAM})--(\ref{SAM}) for arbitrary 3D fields generalize previous analogous results obtained for: (i) nonparaxial free-space fields with well-defined helicity and spin-redirection geometric phase \cite{Bliokh2010}, and (ii) paraxial inhomogeneously-polarized fields and Pancharatnam-Berry phase \cite{Zhang2015}. Our general definitions of phases (\ref{total_phase})--(\ref{geometric_phase}) and (\ref{phase_3DT})--(\ref{Coriolis}) unify these previously unrelated cases. 

\KB{As another instructive example, we consider the relations between the AM and phases in paraxial 2D fields with discrete azimuthal symmetries, such as fields produced by ``q-plates'' \cite{Hasman2005,Marrucci2011,Bliokh2015NP,Rubano2019}. Assuming a q-plate \CMRD{with topological} charge $q$, phase retardation $\delta$, and incident circularly-polarized plane wave with the Jones vector $\left( {\begin{array}{*{20}{c}}
{{1}}\\
{{0}}
\end{array}} \right)$ (helicity $\sigma=1$) or $\left( {\begin{array}{*{20}{c}}
{{0}}\\
{{1}}
\end{array}} \right)$ ($\sigma=-1$) in the circular basis, the output field (\ref{2D_field}) generated by the q-plate can be written as
\begin{equation}
\label{q-plate}
{\bf E}^{(+)}\! =\! \left( {\begin{array}{*{20}{c}}
{{\cos \dfrac{\delta}{2}}}\\
{{i\sin \dfrac{\delta}{2}\, e^{2iq\varphi}}}
\end{array}} \right)\!,~~
{\bf E}^{(-)} \!=\! \left( {\begin{array}{*{20}{c}}
{{i\sin \dfrac{\delta}{2}\, e^{-2iq\varphi}}}\\
{{\cos \dfrac{\delta}{2}}}
\end{array}} \right)\!,
\end{equation}
where the upper index $(\pm)$ indicates the helicity $\sigma$ of the incident wave.} 

\KB{Calculating the total, dynamical, geometric, and C-point phases (\ref{total_phase_2D}), (\ref{dynamical_phase_2D}), (\ref{PB_phase}), and (\ref{C_phase_2D}) along the circular contour $\left\{ {r = {\mathop{\rm const}\nolimits} ,\,\varphi  \in \left( {0,2\pi } \right)} \right\}$ for the fields (\ref{q-plate}), we obtain:
\begin{align}
\label{q-plate_phases}
& \Phi = 2\pi q \sigma (1-\cos\delta), \quad
\Phi_{\rm G} = - 2\pi q \sigma \cos\delta, \nonumber \\
& \Phi_{\rm D} = 2\pi q \sigma, \quad
\Phi_{\rm C} = 2\pi q, 
\end{align}
In turn, the orbital and spin AM densities in the fields (\ref{q-plate}) are given by:
\begin{align}
\label{q-plate_AM}
{\frac{\omega l_z}{w}} = \frac{\Phi}{2\pi}\,, \quad
{\frac{\omega s_z}{w}} = - \frac{\Phi_{\rm G}}{2\pi q}\,.
\end{align}
}
\KB{Thus, the orbital and spin AM of the paraxial q-plate fields can still be associated with the total and geometric phases calculated along the circular contour. However, the total AM $j_z = l_z + s_z$ is generally not quantized and associated with the dynamical phase, because the q-plate field has only discrete azimuthal symmetry. The continuous azimuthal symmetry is present only for $q=1$, when $\omega j_z/w = \Phi_{\rm D}/2\pi = \sigma$.
}

\subsection{\normalsize Extension to electromagnetic fields \\ in optical media}
\label{Sec-Extension}

So far, for the sake of simplicity, we only considered the 3D complex electric field ${\bf E}$ \CKB{as if it were in free space}. The close correspondence between the phases and dynamical properties of the field, revealed in Section~\ref{Sec-AM}, enable us to naturally extend the main definitions of phases to the case of electric and magnetic fields $({\bf E},{\bf H})$ in an isotropic inhomogeneous dispersive lossless medium, characterized by the real-valued permittivity $\varepsilon(\omega,{\bf r})$ and permeability $\mu(\omega,{\bf r})$. 

It is well-known for the energy density \cite{Jackson,Landau_book}, and was recently shown for other dynamical properties of electromagnetic fields, such as momentum and AM \cite{Bliokh2017PRL,Bliokh2017NJP,Alpeggiani2018}, that this extension is realized via the modification of the inner product 
\begin{equation}
\label{inner_product}
{\bf E}^* (...) {\bf E} \to  \frac{1}{2} \left[ \tilde{\varepsilon}\,{\bf E}^*(...) {\bf E} + \tilde{\mu}\,{\bf H}^*(...) {\bf H}\right] ,
\end{equation}
where $(\tilde{\varepsilon},\tilde{\mu}) = (\varepsilon,\mu) + \omega\, d(\varepsilon,\mu)/d\omega$. In particular, for $(...)=\omega$ this yields the well-known Brillouin energy density of \CMRD{the} electromagnetic field. Therefore, it is natural to extend the definitions of phases (\ref{total_phase})--(\ref{geometric_phase}), which are also based on quadratic forms of the field, 
\CMRD{analogously:}
\begin{align}
\label{total_phase_ext}
& \Phi = {\rm Im} \oint \frac{\tilde{\varepsilon}\,{\bf E}^*\! \cdot \left( \nabla  \right){\bf E}+
\tilde{\mu}\,{\bf H}^*\! \cdot \left( \nabla  \right){\bf H}}{\tilde{\varepsilon} \left| {\bf E} \right|^2 + \tilde{\mu} \left| {\bf H} \right|^2} \cdot d{\bf r},\nonumber\\
& {\Phi _{\rm D}} = \frac{1}{2}{{\rm Im}} \oint {\frac{{{\Psi ^*}\left( \nabla  \right)\Psi }}{{{{\left| \Psi  \right|}^2}}} \cdot d{\bf{r}}}
= \pi N_{\rm D}, \nonumber
\\
& {\Phi_{\rm G}} = {\Phi} - {\Phi _{\rm D}},
\end{align}
where 
\begin{equation}
\Psi = \tilde{\varepsilon}\, {\bf E}\cdot {\bf E} + \tilde{\mu}\, {\bf H}\cdot {\bf H} .
\end{equation}
Using these definitions, the relations between the phases and AM (\ref{OAM})--(\ref{SAM}) 
\CMRD{apply} in inhomogeneous dispersive media \cite{Picardi2018}.

\section{\normalsize Conclusions}
\label{Discussion}

\CMRD{In conclusion,} we have described the dynamical and geometric phases, as well as their interplay with polarization singularities, in inhomogeneous polarized fields. We introduced a general formalism (\ref{total_phase})--(\ref{geometric_phase}), which allows determining the total, dynamical, and geometric phases along a spatial contour in a generic multi-component complex field ${\bf E}({\bf r})$. The total phase is determined by the contour integral of the ``local wavevector'' (or normalized canonical momentum density), the dynamical phase is half of the phase calculated in the scalar quadratic field  $\Psi={\bf E}\cdot {\bf E}$, while the geometric phase is the difference between these two phases. In this manner, the dynamical phase is quantized for closed spatial contours as an integer $N_{\rm D}$ times $\pi$. The total phase generically evolves continuously under continuous deformations of the contour, while the dynamical and geometric phases experience opposite $\pi$ jumps when the contour crosses a phase singularity of the field $\Psi$, i.e., a polarization singularity of the vectors field ${\bf E}$. This signifies an important connection between geometric phases and polarization singularities, and, generally, {\it topological} properties of the field.
To the best of our knowledge, this connection was accurately described for the first time in this work.

We have examined thoroughly the two most important and practically-relevant cases, namely 2D and 3D polarizations, which are realized, e.g., in paraxial and nonparaxial optical fields. In a two-component field, the general geometric-phase formula reduces to the well-known Pancharatnam-Berry phase, which plays an important role in polarization optics, liquid-crystal devices and metasurfaces. Importantly, the fundamental form (\ref{PB_phase}) of this phase is not simply equal to the oriented area (solid angle) enclosed by the contour on the Poincar\'e sphere, but \CMRD{acquires} a $\pi N_{\rm C}$ addition, where $N_{\rm C}$ is an integer number characterizing the net topological strength of polarization singularities (C-points) enclosed by the contour in real space, Eqs.~(\ref{C_phase_2D})--(\ref{area}).

In a 3D field, the phase and polarization analysis becomes more sophisticated. The general geometric-phase formula (\ref{geometric_phase_3D}) can be expressed using two very different yet equivalent representations. First, one can interpret geometric phase ``dynamically'' as a Coriolis phase shift (\ref{Coriolis}) induced by the observation of the field with an intrinsic angular momentum (i.e., spin ${\bf S}\propto {\rm Im}({\bf E}^*\!\times {\bf E})$ produced by the field ${\bf E}$) in a locally-rotating reference frame (caused by SO(3) rotations of the polarization ellipse along the contour). This Coriolis treatment unifies the 2D Pancharatnam-Berry phase, when only the spin magnitude (i.e., polarization ellipticity) varies, and the spin-redirection geometric phase for circularly-polarized fields, where only the spin direction (i.e., the orientation of the normal to the polarization ellipse) varies. 

Second, a geometric representation of the polarization evolution and geometric phase is possible on Majorana-type spheres. This representation combines two ${\rm S}^2$ spheres, the Poincar\'e and spin-direction ones, and represents the polarization evolution by two points on a unit sphere in real space, i.e., by four parameters. Furthermore, we have improved the original Majorana-Hannay construction via re-scaling of parameters, to make it better fit the geometric-phase treatment in terms of spherical areas (\ref{Phase_Poincarana}). We call this new representation the ``{\it Poincarana sphere}'', which combines useful features of the Poincar\'e and Majorana spheres. 
 
Most importantly, the Majorana/Poincarana treatment of the geometric phase naturally reveals a remarkable topological role of 3D polarization singularities. Akin to the 2D case, these singularities are C-points in ${\bf r}\in \mathbb{R}^2$ or C-lines in ${\bf r}\in \mathbb{R}^3$. However, in contrast to two-component fields, one cannot characterize the polarization singularities by an integer topological index $N_{\rm C}$. This is because this index counts ``the number of half-turns'' of the polarization ellipse, well-defined for planar SO(2) rotations and meaningless for spatial SO(3) rotations. Nonetheless, there is a meaningful $\mathbb{Z}_2$ topological number (\ref{Z2})characterizing 3D polarization singularities. This number, $M=N_{\rm D}\, {\rm mod}\, 2$, indicates the \CMRD{presence} ($M=1$) or absence ($M=0$) of the {\it polarization M\"obius strip} along the contour of evolution. These two fundamental real-space configurations are represented by one or two contours traced by the representation points on the Majorana/Poincarana sphere. The transition between these two topologically-different cases are realized when the spatial contour crosses a generic C-point. This transition between non-M\"obius and M\"obius cases, is accompanied by: (i) the \CMRD{appearance} of the discontinuity ${\bf A} \to -{\bf A}$ in the polarization-ellipse orientation along the real-space contour, (ii) the reconnection of the two contours on the Majorana/Poincarana sphere into a single joint loop, and (iii) a $\pi$ jump in the geometric phase (\ref{PhG1}) and (\ref{PhG2}).

Thus, the 2D and 3D cases confirm the general conclusion that discontinuous behaviour of the geometric phase near polarization singularities reflects fundamental topological features of multicomponent fields and \CMRD{corresponds} to generic half-integer topological strength of such singularities. 

Finally, we described the close \CMRD{relationship} between the total/dynamical/ geometric phases and the orbital/total/spin angular momenta in cylindrically-symmetric 3D optical fields and indicated the straightforward extension of the general phase formalism to the case of electromagnetic fields in optical media.

We conclude with several remarks on further extensions and implications of the approach described in this work.

1. Throughout this work we considered spatial closed contours for clarity and better visualization of various phases and cases. Note, however, that all the main equations for the total, dynamical, and geometric phases contain the same form $\nabla \cdot d{\bf r}$, which is the differential $d$ of the field along the contour. Therefore, one can apply all these equations to {\it any} evolution of the polarized field, e.g., along the propagation trajectory, in time, etc. This evolution should not be necessarily cyclic, but then the dynamical phase is not quantized anymore and topological numbers (defined only for closed contours) become irrelevant. 

2. It is known that the evolution of two-component complex fields is described by the SU(2) three-parameter group and is conveniently represented on \CMRD{the} ${\rm S}^2$ sphere. Correspondingly, the geometric phase is naturally described on the Poincar\'e sphere using two parameters. In turn, the polarization state of a generic 3D optical field is often described in terms of the {\it eight}-parameter SU(3) group typical for three-level quantum systems \CMRD{\cite{Dennis2004,Carozzi2000,Setala2002,Sheppard2014}}. However, it follows from the Majorana/Poincarana formalism that a fully-polarized 3D  state and the corresponding geometric phase is completely described by only {\it four} parameters, i.e., represented on a \CMRD{symmetrized} double-sphere ${\rm S}^2 \oplus {\rm S}^2 / \, {\mathbb{Z}_2}$. Therefore, one should be careful and distinguish the SU(3) evolution of three-level quantum systems \cite{Khanna1997} from the polarization evolution of spin-1 particles described by a three-component wavefunction \cite{Bouchiat1988}.

3. Geometric phases are ubiquitous in problems involving interference of multi-component fields. These phases have been well described and measured in interference experiments with paraxial optical fields. Our general approach enables one to calculate geometric phases in the evolution or interference of nonparaxial 3D fields, possibly in inhomogeneous media, which are particularly important for nano-optics and photonics. Therefore, one can expect that the general 3D geometric phases will manifest themselves in such problems as, e.g., interference of surface plasmon-polaritons \cite{Zayats2005}, the geometrical-optics quantization of modes of photonic waveguides and cavities \cite{Picardi2018,Ma2016,Kreismann2017}, and propagation/evolution of optical beams in smoothly-inhomogeneous and weakly-anisotropic media \cite{Bliokh2007}. In all these cases, the three components of the field are important. Furthermore, the topological connection between geometric phases and polarization singularities, established in this work, can also manifest itself in various nanooptical problems.

\KB{After this work was submitted, the relevant recent paper \cite{Berry2019} came to our attention. We were pleased to recognize that our general approach to the separation of the dynamical and geometric phases in Sections~\ref{basic_idea} and \ref{Sec-Extension} coincides with that by M. V. Berry and P. Shukla.}

\vspace*{0.3cm}

We are grateful to Ebrahim Karimi and Michael Berry for fruitful discussions. This work was partially supported by the Australian Research Council, the National Science Foundation (PHY-1507278), the Excellence Initiative of Aix Marseille University - A*MIDEX, a French `Investissements d'Avenir' programme, and the Leverhulme Trust Research Programme (RP2013-K-009, SPOCK: Scientific Properties of Complex Knots).

\bibliography{references_BAD_3}

\clearpage

\appendix

\section{Derivation of the Poincarana-sphere construction}
\label{apppoincarana}

In the general Majorana-like construction in Eq.~(\ref{upm}), the parameter $\beta$ is related to the eccentricity of the polarization ellipse, so that for linear polarizations ($|{\bf B}|=|{\bf S}|=0$) $\beta=0$, while for circular polarizations ($\sqrt{2}|{\bf A}|=\sqrt{2}|{\bf B}|=|{\bf S}|=1$) $\beta=1$. The goal is to derive an expression for $\beta$ which provides a straightforward connection between the geometric phase and a solid angle on the Majorana-like sphere. For this purpose, it is useful to express the unit polarization-ellipse vectors $\ua$ and $\ub$ in terms of the unit vectors on the sphere, $\bu_{1,2}$, by using Eq.~(\ref{upm}) and the fact that $\ub=\bar{\bf s}\times\ua$:
\be
\ua=\frac{\bu_1-\bu_2}{2\sqrt{1-\beta^2}},\quad
\ub=\frac{\bu_2\times\bu_1}{2\beta\sqrt{1-\beta^2}}.
\label{abuvecs}
\ee

We now consider the increase of the geometric phase as some parameter $\tau$ is varied (e.g. a parametrization of the spatial contour). 
From Eqs.~(\ref{Coriolis}) and (\ref{angular_velocity}), 
the derivative with respect to $\tau$ of $\Phi_{\rm G}$ can be written as
\be
\deriv\Phi_{\rm G}
=-|{\bf S}|\,\ub\cdot\deriv\ua.
\label{dphiG}
\ee
By substituting Eqs.~(\ref{abuvecs}) into this expression we find
\begin{align}
\label{A3}
\deriv\Phi_{\rm G}&=\frac{|{\bf S}|}{4\beta(1-\beta^2)}(\bu_1 \times \bu_2) \cdot \deriv (\bu_1 - \bu_2) \nonumber\\
=-&\frac{|{\bf S}|\left[\bu_2 \cdot (\bu_1 \times \deriv \bu_1) + \bu_1 \cdot (\bu_2 \times \deriv \bu_2) \right]}{4\beta(1-\beta^2)}.
\end{align}

Consider the first of the two triple products in the last expression, and multiply it by an infinitesimal increment in the parameter, $\tau$:
\begin{align}
\label{triple}
& \delta\tau\,\bu_2 \cdot (\bu_1 \times \deriv \bu_1) 
\nonumber\\
& = \bu_2 (\tau) \cdot \{ \bu_1(\tau) \times [\bu_1(\tau)+\delta\tau\,\deriv \bu_1(\tau)]\} \nonumber\\
& =\bu_2(\tau) \cdot \bu_1(\tau)\times\bu_1(\tau+\delta\tau).
\end{align}
%
It is shown at the end of this appendix that the triple product in the last expression of (\ref{triple}) is related to the oriented area (i.e. solid angle) $\delta\Sigma_{\rm I}$ of the geodesic triangle on the sphere formed by the three points in this expression as
\begin{align}
& \bu_2(\tau) \cdot [\bu_1(\tau) \times \bu_1(\tau+\delta\tau)] 
\nonumber\\
& =[1+\bu_1(\tau) \cdot \bu_2(\tau)]\,\delta\Sigma_{\rm I}
=2\beta^2\,\delta\Sigma_{\rm I}.
\end{align}
However, what turns out to be more useful is the oriented area 
$\delta\tilde{\Sigma}_{\rm I}$ formed by the spherical triangle with corners $\bu_1(\tau+\delta\tau)$, $\bu_1(\tau)$, and $-\bu_2(\tau)$, so we use instead
\begin{align}
&-\bu_2(\tau) \cdot \left[ \bu_1(\tau) \times \bu_1(\tau+\delta\tau)\right] 
\nonumber\\
&=-[1-\bu_1(\tau) \cdot \bu_2(\tau)] \,\delta\tilde{\Sigma}_{\rm I}
=-2(1-\beta^2)\,\delta\tilde{\Sigma}_{\rm I}.
\end{align}

Following similar steps with the second triple product in Eq.~(\ref{A3}), we find:
\be
\label{A7}
\delta\Phi_{\rm G} = \deriv\Phi_{\rm G}\delta\tau = \frac{|{\bf S}|}{2\beta}(\delta\tilde{\Sigma}_{\rm I}+ \delta\tilde{\Sigma}_{\rm II}) = \frac{|{\bf S}|}{2\beta}\,\delta{\Sigma}_{\rm geod},
\ee
where $\delta\tilde{\Sigma}_{\rm II}$ is the area of the spherical geodesic triangle with corners $\bu_1(\tau)$, $-\bu_2(\tau)$, and $-\bu_2 (\tau+\delta\tau)$, or equivalently $\bu_1(\tau+\delta\tau)$, $-\bu_2(\tau)$, and $-\bu_2(\tau+\delta\tau)$. Adding the two infinitesimal triangles, we find that the area $\delta{\Sigma}_{\rm geod}$ in Eq.~(\ref{A7}) is the area of the polygon with four corners $\bu_1(\tau+\delta\tau)$, $\bu_1(\tau)$, $-\bu_2(\tau)$, and $-\bu_2(\tau+\delta\tau)$, i.e., the {\it oriented area swept by the geodesic segment joining $\bu_1$ and $-\bu_2$} (or equivalently $-\bu_1$ and $\bu_2$), see Fig.~\ref{Lambert}.

For the geometric phase to have an interpretation that is simply proportional to this area, as on the Poincar\'e sphere, $\beta=|{\bf S}|$ is the most natural choice. Then, Eq.~(\ref{A7}) yields $\Phi_{\rm G}= {\Sigma}_{\rm geod}/2$,  Eq.~(\ref{Phase_Poincarana}), after integration along the contour.

{\bf Infinitesimal solid angle element.} Consider three unit vectors $\bu$, $\bv$ and $\bv'$, the latter two being infinitesimally close. Their triple product gives
\be
\bu\times\bv\cdot\bv'=\bu\cdot\bv\times\bv'=\sin\theta\,\bw\cdot\bv'=\sin\theta\,(\phi\,\sin\theta),
\ee
where $\theta$ is the angle between $\bu$ and $\bv$, $\bw$ is a unit vector normal to $\bu$ and $\bv$, and $\phi$ is the infinitesimal angle between $\bv'$ and the plane containing $\bu$ and $\bv$ (Fig.~\ref{figapp}).
The area of the geodesic triangle formed by the three vectors (shown in green in Fig.~\ref{figapp}) is easiest to calculate by using a spherical coordinate system whose polar axis is $\bu$, so that $\phi$ is simply the azimuthal separation between $\bv$ and $\bv'$. This area is then $\Sigma^+=(1-\cos\theta)\,\phi$. Similarly, the area of the complement of this wedge, enclosed by $-\bu$, $\bv$ and $\bv'$, is $\Sigma^-=(1+\cos\theta)\,\phi$. The relation of these areas/solid angles and the triple product is then
\begin{align}
\Sigma^{\pm}&=(1\mp\cos\theta)\,\phi=(1\mp\cos\theta)\frac{\bu\times\bv\cdot\bv'}{\sin^2\theta}\nonumber\\
&=\frac{\bu\times\bv\cdot\bv'}{1\pm\cos\theta}=\frac{\bu\times\bv\cdot\bv'}{1\pm\bu\cdot\bv}.
\end{align}

\begin{figure}
\centering
\includegraphics[width=.5\linewidth]{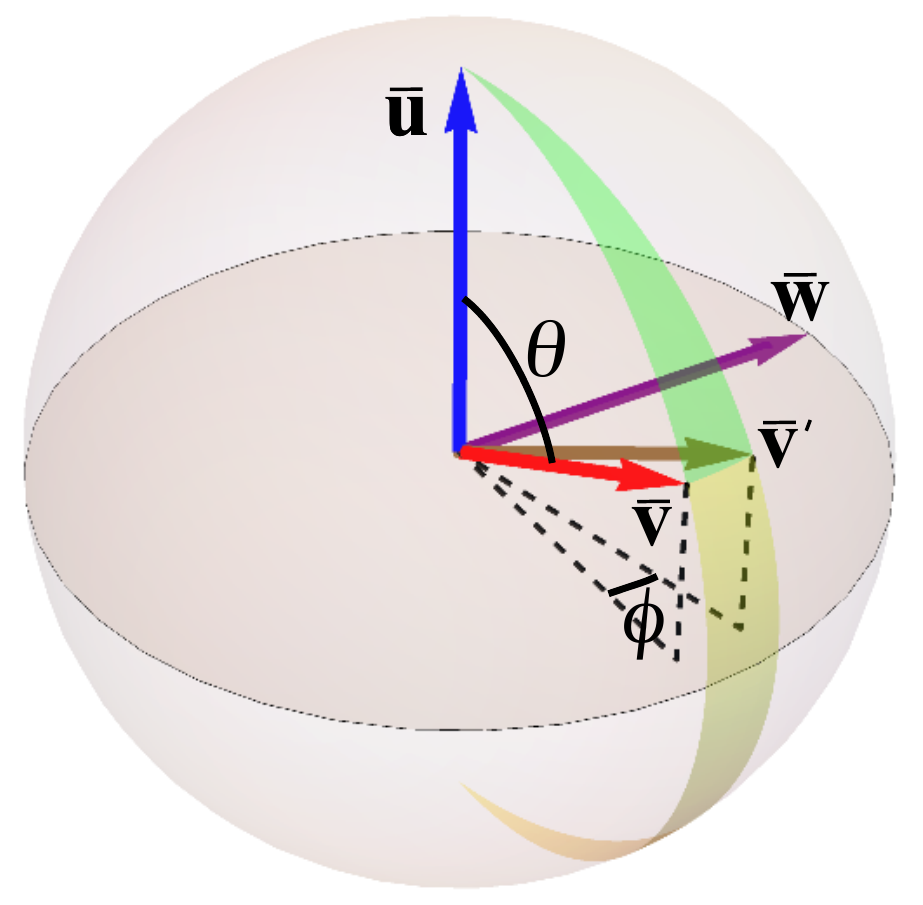}
\caption{The green area equals the solid angle enclosed by the geodesic triangle formed by $\bu$ (blue arrow), $\bv$ (red arrow), and $\bv'$ (brown arrow). The corresponding solid angle enclosed by $-\bu$, $\bv$ and $\bv'$ is shown as a yellow area. Also shown are the vector $\bw$ (purple arrow), the angles $\theta$ between $\bu$ and $\bv$, and the infinitesimal angle $\phi$ between the projections of $\bv$ and $\bv'$ onto the plane normal to $\bu$. Note $\bv$ and $\bv'$ are shown as being at the same angle $\theta$ from $\bu$. If the infinitesimal line joining $\bv$ and $\bv'$ where not normal to $\bu$, this would cause only a change in the solid angle of the order of $\phi^2$, which can be ignored.}
\label{figapp}
\end{figure}

\end{document}